\documentclass[twocolumn,preprintnumbers,amsmath,amssymb]{revtex4}
\usepackage{color}
\usepackage{bm, graphicx, amsmath}
\usepackage{hyperref}

\begin{document}
	
	\title{Demonstrating NISQ Era Challenges in Algorithm Design on IBM's 20 Qubit Quantum Computer}
	\author{Daniel Koch$^{1}$, Brett Martin$^{2}$, Saahil Patel$^{1}$, Laura Wessing$^{1}$, Paul M. Alsing$^{1}$}
	\affiliation{$^{1}$Air Force Research Lab, Information Directorate, Rome, NY}
	\affiliation{$^{2}$Air Force Academy, Colorado Springs, Co}
	
	\begin{abstract}
		
		As superconducting qubits continue to advance technologically, the realization of quantum algorithms from theoretical abstraction to physical implementation requires knowledge of both quantum circuit construction as well as hardware limitations.  In this study we present results from experiments run on IBM's 20-qubit `Poughkeepsie' architecture, with the goal of demonstrating various qubit qualities and challenges that arise in designing quantum algorithms.  These include experimentally measuring $T_1$ and $T_2$ coherence times, gate fidelities, sequential CNOT gates, techniques for handling ancilla qubits, and finally CCNOT and QFT$^{\dagger}$ circuits implemented on several different qubit geometries.  Our results demonstrate various techniques for improving quantum circuits which must compensate for limited connectivity, either through the use of SWAP gates or additional ancilla qubits.

	\end{abstract}
	
	\maketitle
	
	\section{Introduction}
	
	For as long as technology remains in the NISQ (Noisy Intermediate Scale Quantum) Era \cite{nisq} of quantum computers, quantum algorithm design will need to compensate for noisy qubits.  While algorithms such as Shor's \cite{shor} and Grover's \cite{grover} have proven mathematical speedups over the best known classical algorithms, they critically rely on demands from quantum computers such as qubit coherence times and gate fidelities, which are to date unfeasible.  For superconducting qubits specifically, these various sources of noise \cite{noise1,noise2,noise3,noise4} inhibit the success of quantum algorithms, which in turn diminish or completely negate their potential for speedups. Even still, technological strides and new techniques for minimizing noise continue to develop \cite{err1,err2,err3}, with the hope that someday soon we will reach full error-correcting \cite{err_corr1,err_corr2,err_corr3} quantum computers.
	
	Due to the complex technological nature of quantum computers, the current standard model by which interested users can work with these machines is through remote access with various vendors \cite{ibmq, google, rigetti, microsoft}.  Analogous to high-level classical programming languages, these vendors offer quantum programming languages which grant the ability for users to execute quantum circuits, without necessarily knowing the full technical extent of how they are implemented via superconducting qubits.  Consequently, this allows for an important separation of quantum software from hardware, opening up more opportunities for research efforts in the field of quantum algorithms\cite{tutorials1,tutorials2}.  In the spirit of this new dawn of quantum programming, the findings in this study reflect the capabilities and limitations of this current model for quantum computer access, aiming to test the 20-qubit Poughkeepsie architecture through various experiments.
	
	Each experiment in this study is motivated by different components which are critical to the success of larger, more complex quantum algorithms.  These include $T_1$ and $T_2$ coherence times \cite{T2_1,T2_2, T1_T2}, single and 2-qubit gate fidelities, and qubit connectivity.  After testing these properties individually, we then  study their combined effects through implementations of CCNOT and QFT$^{\dagger}$ circuits \cite{toffoli,qft}.  Throughout these various experiments, we make a concentrated effort to distinguish between results which are simply technological benchmarks (coherence times, gate fidelity, etc.) and those which are more fundamental to algorithm design. Our findings demonstrate several challenges which must be factored into NISQ Era algorithm design, adding to the growing population of studies which aim to benchmark IBM's qubits \cite{ibmq1,ibmq2,ibmq3}, as well as test the limits of various algorithm implementations \cite{ibm_exp1,ibm_exp2,ibm_exp3,ibm_exp4,ibm_exp5}.

	\subsection{Layout}
	
	The layout of this paper is as follows: In section 2 we investigate the $T_1$ and $T_2$ coherence times of various qubits.  In section 3 we demonstrate CNOT gate fidelities across all 20 of IBM's Poughkeepsie qubits, showing the extent to which a single CNOT operation can be reliably performed between distant qubits.  Section 4 contains no experimental results, but lays the framework and motivation for the remainder of the study.  In sections 5 and 6 we experimentally implement CCNOT and QFT$^{\dagger}$ circuits on various qubit geometries.  And lastly, section 7 summarizes the main results of the paper and their implications for future algorithm design.
	
	\section{Coherence Times}
	
	In quantum computing, a qubit is a two-level system that can simultaneously occupy both the $|0\rangle$ (ground) and $|1\rangle$ (excited) states through superposition, and is also sensitive to the relative phase between the two states.  In practice however, one must always be mindful of the potential for noise to cause qubits to deviate from their intended states.  Contrary to their classical counterparts, current qubits have short timescales for which their quantum states are usable in any sort of calculation.  These time frames are referred to as coherence times, quantified by the metrics $T_1$ and $T_2$, representing timescales after which a qubit has likely lost its computational utility.  Physically, these metrics correspond to a qubit's interactions with a noisy environment, tracking the probability that a qubit's excited state ($T_1$) or superposition state ($T_2$) is preserved, which both decay exponentially in time.  Equation \ref{Eqn:Exp_Decay} below shows the probability of a qubit resisting a decoherence collapse after an interval of time $\Delta t$.
	
	\begin{eqnarray}
	\textrm{P}_{i}(\Delta t) = e^{\frac{- \Delta t}{T_i}}
	\label{Eqn:Exp_Decay}
	\end{eqnarray}
	
	While working on IBM's Poughkeepsie architecture, coherence times for both $T_1$ and $T_2$ ranged from lows of $30$ - $40 \mu s$, to highs of $100$ - $120 \mu s$.   In this section we present experimental results aimed to verify these coherence times through direct observations of decoherence over varying timescales.  However, when working on these shared devices remotely, one must be mindful of other users, which can cause certain times during the day to be more competitive for machine usage than others.  This in turn can be problematic for experiments which require data collecting from numerous trials, such as the coherence experiments to come, as we found that qubit coherence times can fluctuate throughout a 24 hour day.  Thus, the results shown in the coming two subsections represent some of the best data obtained as a remote user, overcoming the challenge of shared device usage and ultimately demonstrating the coherence times of IBM's qubits.
	
	\subsection{T$_1$ Energy Relaxation}
	
	The $T_1$ metric corresponds to a spontaneous decay from the excited state ($\hspace{.05cm}|1\rangle \hspace{.01cm}$) to the ground state ($\hspace{.05cm}|0\rangle \hspace{.01cm}$).  	Just as classical computing is reliant on the long shelf life of bits, a critical ingredient for quantum circuits is how long a qubit can maintain the $|1\rangle$ state.  2-qubit gates such as CNOT and control-$R_{\phi}$, which make up the backbone of several critical quantum subroutines, are reliant on `control' qubits whereby the action of the 2-qubit gate is only performed if the control qubit is in the $|1\rangle$ state.  Thus, the impact of a spontaneous energy relaxation in the middle of an algorithm can vary depending on when, and on which qubit the error occurs.  If a key qubit to a circuit's success were to unintentionally undergo a $T_1$ collapse, it could spell the end of the algorithm.  Conversely, as demonstrated in some of the later experiments, certain algorithms can still yield successful results despite one or more qubits undergoing spontaneous decays, provided the collapse happens after a qubit has served its purpose.
	
In order to experimentally demonstrate the decay shown in equation \ref{Eqn:Exp_Decay}, figure \ref{Fig:T1_Circuit} below shows the circuit used to verify the underlying $T_1$ nature of IBM's qubits.  The circuit is designed such that the qubit is initially brought into the $|1\rangle$ excited state via an $X$ gate, followed by a desired amount of time $\Delta t$ whereby we anticipate an energy relaxation according to the exponential probability distribution.
	
	\begin{figure}[h]                     
		\centering
		\includegraphics[scale=.5]{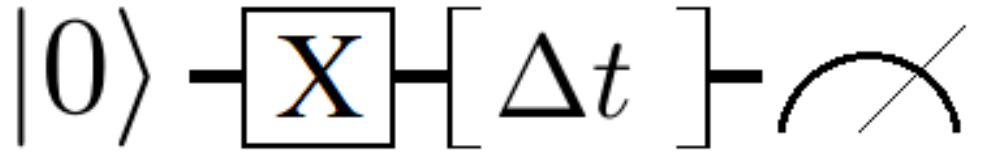}
		\caption{Quantum circuit for studying $T_1$ coherence times.  The qubit is excited into the $|1\rangle$ state via the $X$ gate, followed by various amounts of time $\Delta t$ where the qubit may spontaneously undergo a $T_1$ collapse.}
		\label{Fig:T1_Circuit}
	\end{figure}
	
\begin{figure}[h]                     
		\centering
		\includegraphics[scale=.29]{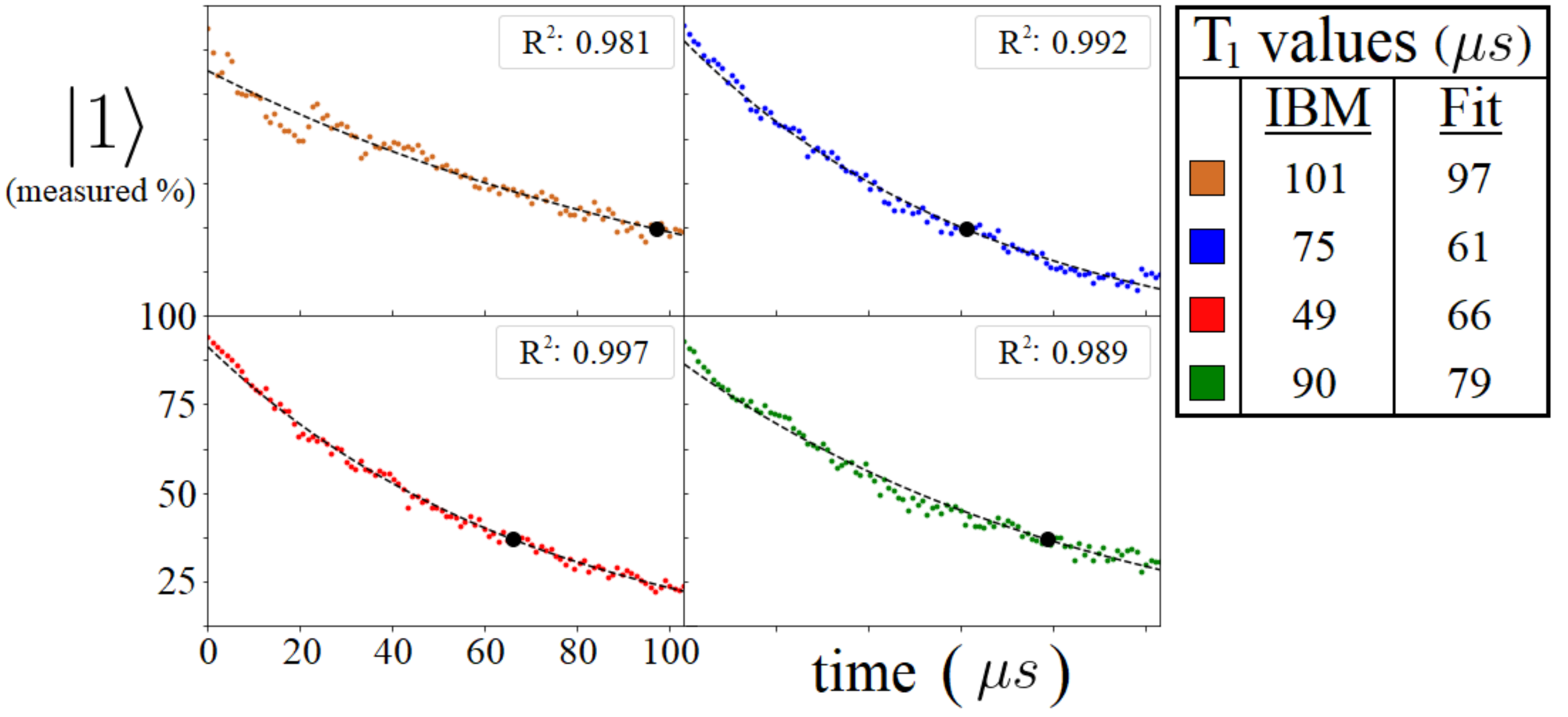}
		\caption{ (scatter plots) Data collected running the circuit shown in figure \ref{Fig:T1_Circuit} for various qubits on IBM's Poughkeepsie architecture.  Accompanying each set of data are exponential regression best fits (dashed lines) used to extract experimental $T_1$ values (black circles), along with their associated correlation coefficients R$^2$.  These $T_1$ values are shown in the accompanying table, as well as the reported times from IBM.}
		\label{Plt:T1_Fit}
	\end{figure}

	In performing the experiment in figure \ref{Fig:T1_Circuit}, various $\Delta t$ times were tested in order to reveal the full exponential decaying nature of the qubits.  For each value of $\Delta t$ the circuit was run 8000 times, from which the results were then used to compute an average percentage probability of decay.  Once all of the experiments for a given qubit were completed, exponential regression fits were then performed to the data. The $T_1$ values from these best fits are displayed alongside the plots in figure \ref{Plt:T1_Fit}, as well as the reported $T_1$ times by IBM for each qubit.
	
	\subsection{T$_2$ Transverse Relaxation}
	
	By comparison to $T_1$, which has a single well defined physical description, $T_2$ coherence for qubits can take on several potential definitions.  In this study, we present results based on two experiments, commonly referred to as `$T_2$ Ramsey' and `$T_2$ Echo'.  The quantum circuits for each experiment are shown below in figure \ref{Fig: T2 Circuits}.  In both experiments the qubit is initially brought into the 50-50 superposition state $|+\rangle$ via a Hadamard gate, followed by various amounts of time $\Delta t$, and finally a second Hadamard just before the measurement.  During the time between Hadamard gates, the qubit is subject to spontaneous energy relaxation ($T_1$) as well as dephasing and frequency drifting, for a combined effect referred to as transverse relaxation \cite{noise3}.
	
	\begin{figure}[h]                     
		\centering
		\includegraphics[scale=.35]{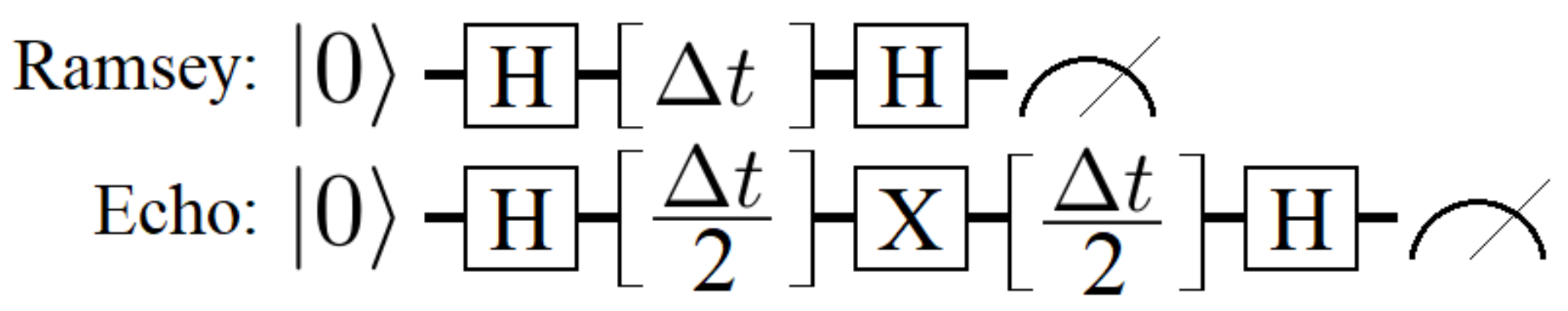}
		\caption{Quantum circuits for demonstrating the $T_2$ nature of IBM's qubits.  The difference between the two experiments can be seen in the extra $X$ gate which splits the time $\Delta t$, which is used to counteract the drifting superposition state.}
		\label{Fig: T2 Circuits}
	\end{figure}
	
\begin{figure}[h]                     
		\centering
		\includegraphics[scale=.28]{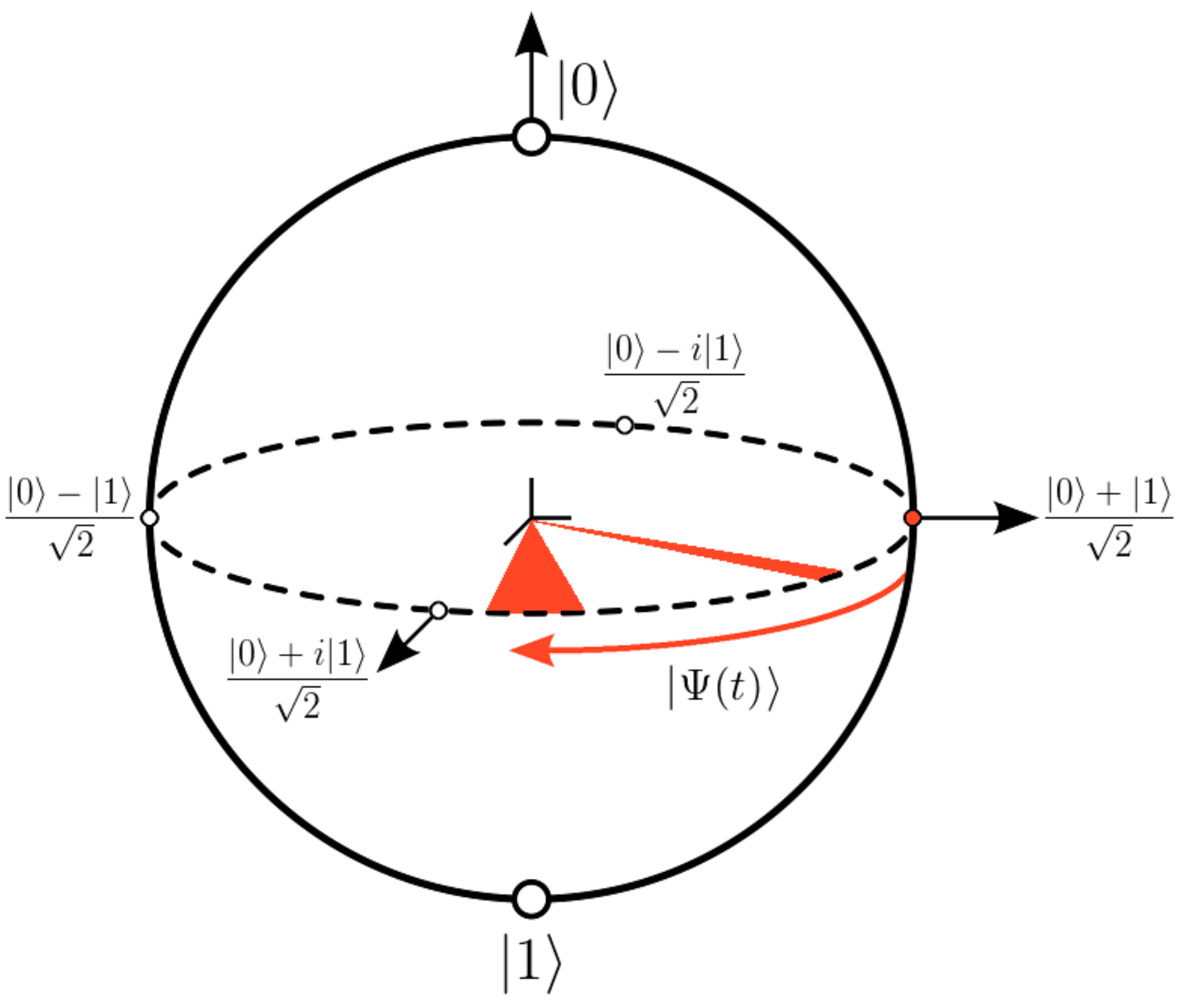}
		\caption{Bloch Sphere representation of the state of the qubit after being initialized by a Hadamard gate.  The orange shaded areas in the equatorial plane represent the growing uncertainty of the state of the qubit as it drifts, reaching a fully decoherent state after enough time.}
		\label{Fig: T2 Bloch}
	\end{figure}

	Beginning with the Ramsey experiment, the theoretical final state after two sequential Hadamard gates ($\Delta t = 0$) should return the qubit back to the ground state.  However, when time is introduced in between these two $H$ gates, the qubit becomes susceptible to $T_2$ transverse relaxation.  Illustrated in figure \ref{Fig: T2 Bloch}, the frequency drifting component of $T_2$ relaxation causes the state of the qubit to process around the equatorial plane of the Bloch Sphere.  

	\begin{eqnarray}
	|\Psi (t) \rangle \hspace{.15cm} = \hspace{.15cm} \frac{|0\rangle \hspace{.05cm}+\hspace{.05cm}e^{i \omega t}|1\rangle }{\sqrt{2}}
	\label{Eqn: T2 Drift}
	\end{eqnarray}	
	
	Experimentally, this effect causes the second Hadamard gate to transform the qubit to a new final state based on the elapsed time, one which oscillates between $|0\rangle$ and $|1\rangle$ with the frequency of the drift.  Secondly, the qubit is also subject to pure dephasing over time, represented by the growing shaded area in figure \ref{Fig: T2 Bloch}, whereby one gradually loses knowledge of the exact state of the qubit \cite{book1}.
	
	The frequency of the precessing state shown above in equation \ref{Eqn: T2 Drift} can be determined through the difference in energies of the two-level qubit system, $\omega = E_1 - E_0$.  As a result, the probability of measuring the $|0\rangle$ state in the Ramsey experiment as a function of time goes like cos$^2$(), which can be seen in figure \ref{Plt: T2 Ramsey} below.
	
		\begin{eqnarray}
	\textrm{P}( \hspace{.04cm}  |0\rangle \hspace{.03cm} ) \hspace{.1cm}=\hspace{.1cm} |\hspace{.03cm}\langle 0 | \textrm{H} | \Psi(t) \rangle \hspace{.03cm} | ^2 \hspace{.1cm} = \hspace{.1cm} \textrm{cos}^2\Big{(} \hspace{.02cm} \frac{\omega t}{2} \hspace{.02cm} \Big{)}
	\label{Eqn: T2 |0> Probability}
	\end{eqnarray}
	
	In order to obtain data similar to that of the $T_1$ experiment, suitable for an exponential best fit, one can counteract the drifting process of the qubit using the $T_2$ Echo circuit.  By splitting each time evolution ($\Delta t$) into two equal halves and using an additional $X$ gate, one can refocus the qubit back to the $|+\rangle$ state just prior to the second Hadamard, guaranteeing a theoretical measurement of $|0\rangle$.  Visually, one can picture this process as the state of the qubit drifting along for some time $\Delta t$ (figure \ref{Fig: T2 Bloch}), becoming $|\Psi (\Delta t) \rangle$ (equation \ref{Eqn: T2 Drift}), undergoing a reflection as a result of the $X$ gate, and finally drifting once again of equal time back to its starting state.
	
\begin{figure}[h]                     
		\centering
		\includegraphics[scale=.35]{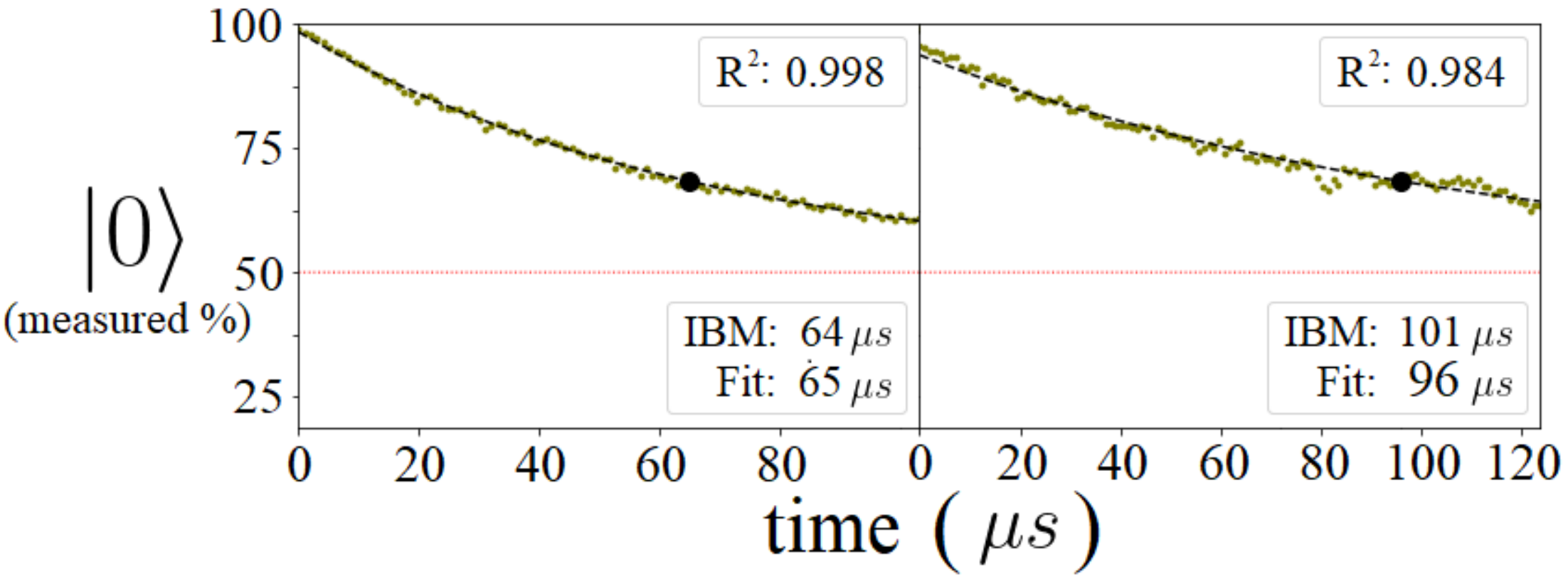}
		\caption{(green circles) Data collected running the $T_2$ Echo circuit for two different qubits. (dashed line) Exponential best fits used to extract experimental values for $T_2$ (black circle) to compare with values reported by IBM.}
		\label{Plt: T2 Echo}
	\end{figure}

	Plotted below are the results obtained running the Ramsey and Echo experiments on various qubits on the Poughkeepsie architecture.  Figure \ref{Plt: T2 Ramsey} shows a typical Ramsey experiment, whereby the probability of the final state oscillates between $|0\rangle$ and $|1\rangle$ as a function of time, while simultaneously dampening into a fully decohered state as a result of dephasing.  Figure \ref{Plt: T2 Echo} illustrates the Echo technique described earlier, showing the effect of using an $X$ gate to let the quantum system naturally refocus the state of the qubit.  Exponential best fits for the $T_2$ Echo experiment are given, along with reported values from IBM.
	
	\begin{figure}[h]                     
		\centering
		\includegraphics[scale=.35]{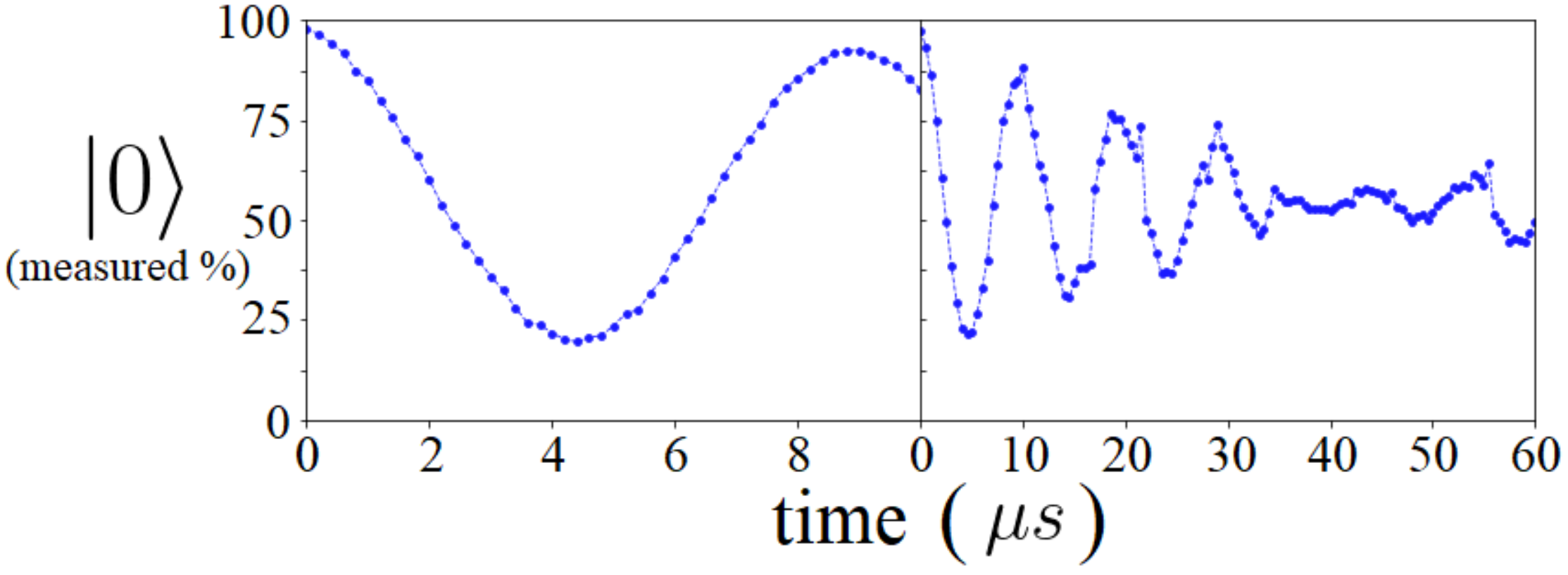}
		\caption{Data collected running the $T_2$ Ramsey circuit \ref{Fig: T2 Circuits} for two different time scales, both on the same qubit.}
		\label{Plt: T2 Ramsey}
	\end{figure}
	
	\section{Chaining CNOT Gates}
	
	When comparing quantum algorithms to classical competitors, claims of speedups often assume full connectivity between all qubits in the quantum system.  By connectivity, we refer to the ability for two qubits to perform a 2-qubit operation.  If one looks to the foreseeable future of qubit technologies however, it is possible that a fully connected superconducting 20 or 50+ qubit device may be upwards of a decade or more away.  Thus, in order to compensate for lacking connectivity, quantum algorithms will need to be adapted to fit the various existing architectures.
	
\begin{figure}[h]                     
		\centering
		\includegraphics[scale=.2]{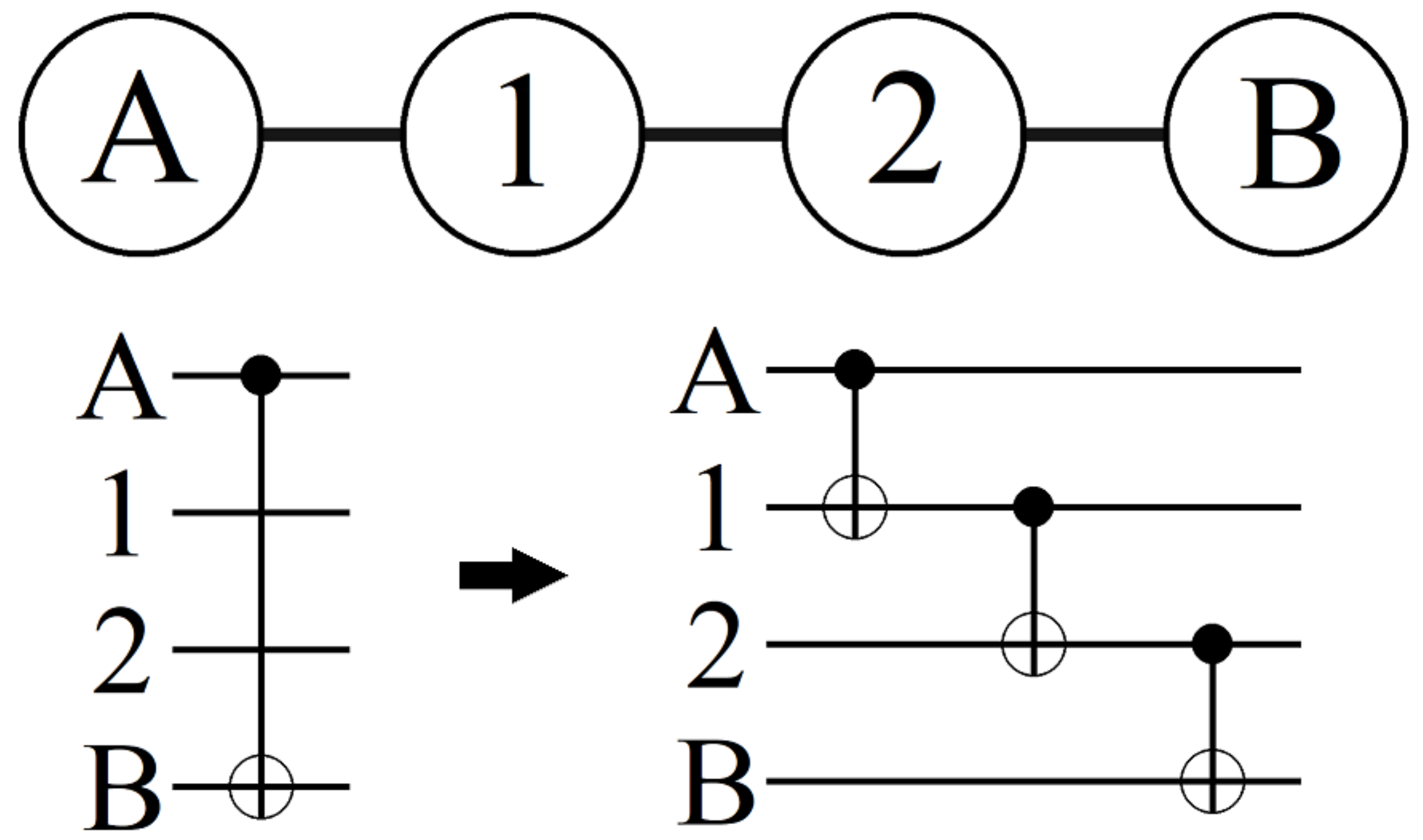}
		\caption{An example of a CNOT gate implementation between distant qubits $A$ and $B$, which lack a direct connection.  Qubits $1$ and $2$ serve as ancilla, acting as intermediate control qubits in order to pass along the desired effect from qubit $A$. }
		\label{Fig:CNOT Chain}
	\end{figure}

	In this section we investigate the effectiveness of using CNOT gates (control-$X$ gates) as a means of compensating for limited qubit connectivity.  We study the reliability with which one can use a series of CNOT gates to invoke a control operation between distant qubits, which do not directly share a connection.  Figure \ref{Fig:CNOT Chain} shows an example of a length-3 chain (two intermediate qubits separating the control and target), achieving a CNOT operation between qubits $A$ and $B$.
	
	In figure \ref{Fig:CNOT Chain}, qubits such as $1$ and $2$, which only serve an intermediate means for connecting $A$ and $B$, are often referred to as ancilla qubits.  Such qubits play a pivotal role in delivering the control operation between the distant computational qubits ($A$ and $B$), and in principle are meant to have no direct impact on the success of the algorithm.  In practice however, this last point can be difficult to control, as merely their incorporated presence in the quantum system can lead to new sources of error.  Additionally, this problem can become compounding as certain algorithms may require the use of the same ancilla qubits several times, requiring proper handling of these qubits after each use.
	
	\subsection{Delivering Desired States}
	
\begin{figure}[h]                     
		\centering
		\includegraphics[scale=0.3]{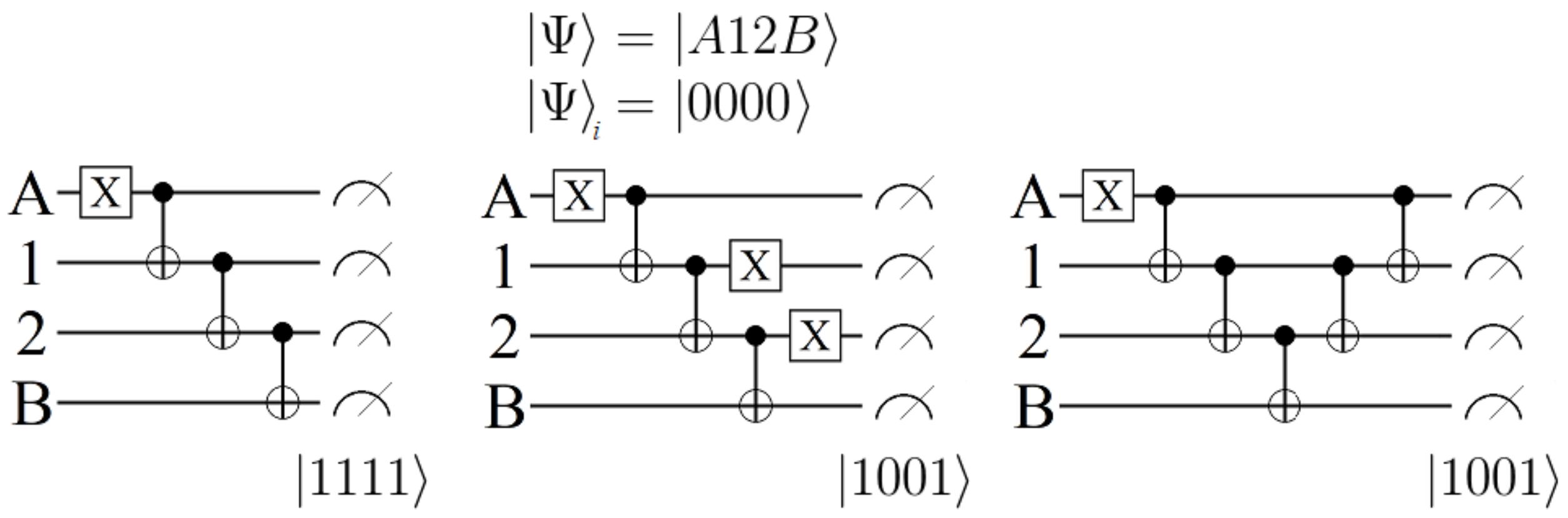}
		\caption{(left) CNOT chain circuit where the desired final state of the system leaves all of the ancilla qubits in the $|1\rangle$ state. (center) Modified version of the left circuit, where the presence of additional $X$ gates now result in the desired final state of each ancilla qubit to be $|0\rangle$. (right) The most general form for a CNOT chain, using additional CNOT gates to properly reset each ancilla qubit when the starting control qubit is in a superposition state.}
		\label{Fig:Chain Circuits}
	\end{figure}

	In testing the effectiveness of IBM's 20-qubit device for implementing CNOT chains, the overall goal of each experiment is to measure both the starting and final (control and target) qubits in the $|1\rangle$ state. This is done by exciting the initial control qubit into the $|1\rangle$ state with an $X$ gate, followed by a series of CNOT gates along some path of ancilla qubits.  For each path studied we examine three circuits (figure \ref{Fig:Chain Circuits}), two such that the desired final state of the system contains all of the ancilla qubits in the $|0\rangle$ state, and one for $|1\rangle$.  The motivation for studying multiple variations of each chain circuit stems from whether or not a certain algorithm requires the ancilla qubits to be reset for future use, oftentimes determined by whether or not the control qubit contains superposition. 
	
	The left circuit in figure \ref{Fig:Chain Circuits} is the simplest form of a CNOT chain, where the ideal final state of the system leaves each ancilla qubit in the $|1\rangle$ state.  More generally, this circuit represents the case where the ancilla qubit states are inconsequential.  That is to say, the only desired effect is such that the distant control and target qubits are both in the $|1\rangle$ state, completing the effect of the computational CNOT gate between them with no intended future use of the ancilla qubits.  Conversely, circuits of the center and right type are designed such that the desired final state of the system leaves each ancilla qubit back in the $|0\rangle$ state, representing the case where one anticipates future use from these ancilla.  For instances where one knows that the initial control qubit is purely in the $|1\rangle$ state, the central circuit would be optimal due to the use of only single qubit gates for resetting the ancilla.  However, for more general cases in which the control qubit may contain superposition, the additional CNOT gates are necessary for resetting.
	
	\subsection{Experimental Design}
	
	In designing CNOT chain paths for IBM's 20-qubit Poughkeepsie architecture, figure \ref{Fig:Chain Path} below illustrates the general layout for each experiment, showcasing the longest path of ancilla qubits tested for a single CNOT chain (touching all 20 qubits on the device).  In addition to the maximum, all intermediate lengths were tested as well, keeping the control qubit fixed and moving the final target qubit along the path shown.
	
	\begin{figure}[h]                     
		\centering
		\includegraphics[scale=.18]{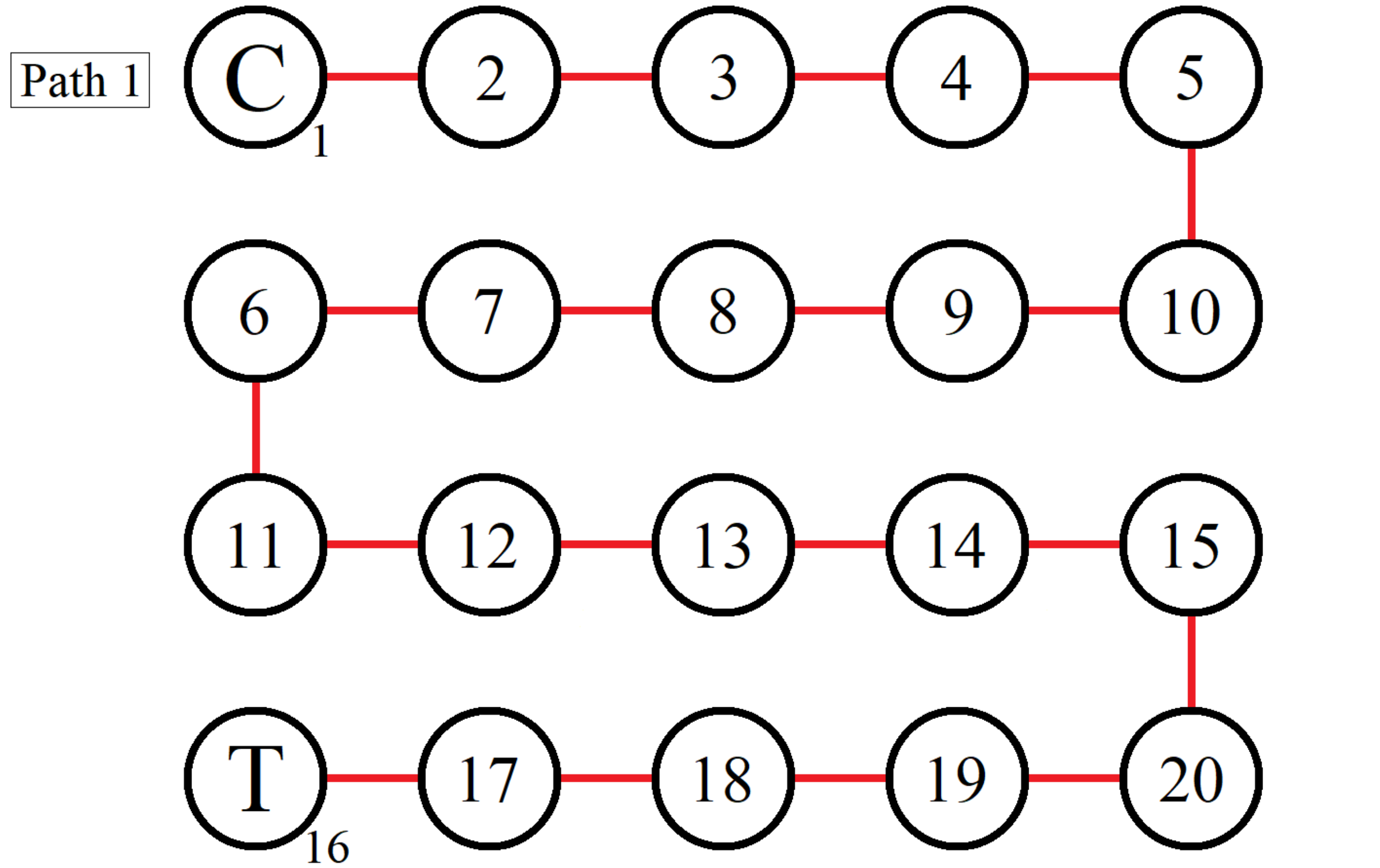}
		\caption{Example of a full CNOT chain path on the Poughkeepsie architecture.  One experimental run studies the success rates for implementing a CNOT operation between the starting control qubit (C) and the final target qubit (T), tested for all chain lengths from $1$ to $19$.}
		\label{Fig:Chain Path}
	\end{figure}
	
\begin{figure}[h]                     
		\centering
		\includegraphics[scale=.28]{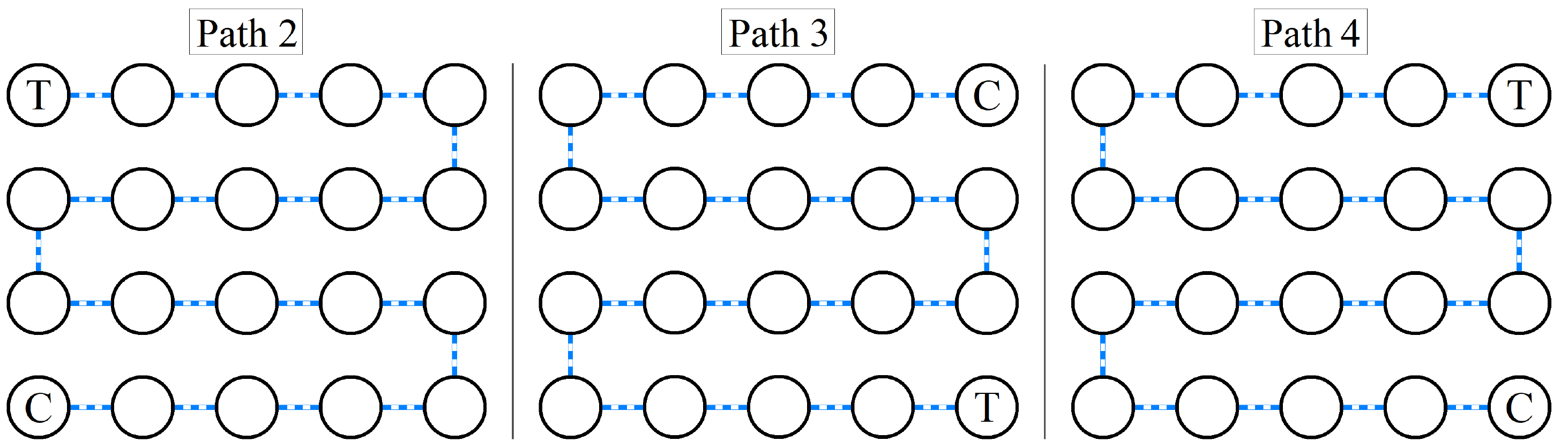}
		\caption{Additional orientations for the full 20-qubit CNOT chain experiment.  Individual results for each orientation can be seen in figure \ref{Plt: Paths Compared}, as well as the combined average fidelity rates in figure \ref{Fig:Chain Circuits}}
		\label{Fig:Chain Path Alts}
	\end{figure}

	The path shown in figure \ref{Fig:Chain Path} is one of four orientations tested experimentally.  In order to achieve the best average result for CNOT chain success, as well as potentially identify any trends for certain qubits, three additional orientations were also tested, shown in figure \ref{Fig:Chain Path Alts}.
	
	\subsection{Experimental Results}
	
	Presented here are the results experimentally gathered for the various CNOT chain experiments, testing each of the circuit types illustrated in figure \ref{Fig:Chain Circuits} across all four path orientations.  In all of the coming results, we distinguish the measurement outcomes from each experiment into three categories based on the state of the control, target, and ancilla qubits.   Firstly, any measurement outcome yielding either the control or target qubit in the $|0\rangle$ state is considered a failure.   Next, we separate cases whereby the CNOT chain successfully yields both the control and target qubits in the $|1\rangle$ state into two groupings based on the final state of the ancilla qubits.   The more lenient of the two metrics, which we call $f_1$, tracks the final state of only the control and target qubits, regardless of the ancilla qubits.  Conversely, the second state fidelity metric, $f_2$, tracks the percentage of trials where $\textit{all}$ qubits in the system are found to be in their theoretically desired state of either $|0\rangle$ or $|1\rangle$.
	
	\begin{eqnarray}
	f_1 \hspace{.15cm} &\equiv&  \hspace{.15cm}  \big{|}\hspace{.05cm} \langle \hspace{.02cm}11\hspace{.02cm}|\hspace{.02cm}\textrm{CT}\hspace{.02cm}\rangle \hspace{.05cm} \big{|}^2 \\
	f_2 \hspace{.15cm} &\equiv& \hspace{.15cm}  \big{|}\hspace{.05cm}   \langle \hspace{.02cm}11\hspace{.02cm}|\hspace{.02cm}\textrm{CT}\hspace{.02cm} \rangle \otimes \langle \hspace{.02cm}\textrm{A}'\hspace{.02cm}|\hspace{.02cm}\textrm{A}\hspace{.02cm}\rangle^{\otimes N}  \hspace{.05cm} \big{|}^2
	\label{Eqn: CNOT Chain Fidelities}
	\end{eqnarray}
	
\begin{figure}[h]                     
		\centering
		\includegraphics[scale=.55]{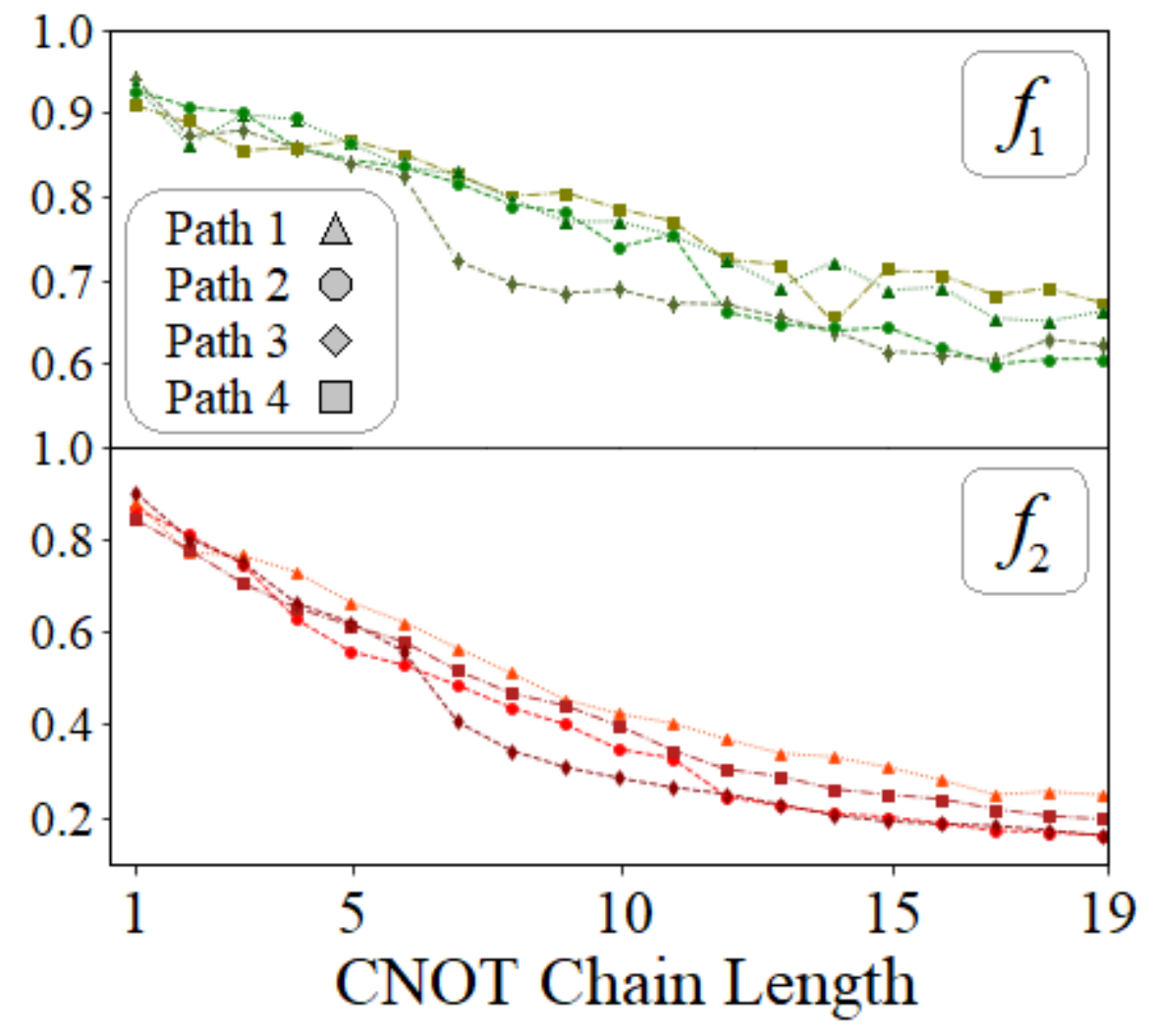}
		\caption{Comparison of the four CNOT chain path orientations shown in figures \ref{Fig:Chain Path} and \ref{Fig:Chain Path Alts}, for the case where the desired state of each ancilla qubits is $|0\rangle$ using $X$ gates (middle circuit in figure \ref{Fig:Chain Circuits}). }
		\label{Plt: Paths Compared}
	\end{figure}

	In the equations shown above, the states $|\hspace{.02cm}\textrm{CT}\hspace{.02cm}\rangle$ and $|\hspace{.02cm}\textrm{A}\hspace{.02cm}\rangle$ represent the final measured states of the control, target, and intermediate ancilla qubits respectively in the computational basis.  The state $|\hspace{.02cm}\textrm{A}'\hspace{.02cm}\rangle$ represents the desired final state for the ancilla qubits, either $|0\rangle$ or $|1\rangle$ according to the circuit types laid out in figure \ref{Fig:Chain Circuits}.  Using the metrics $f_1$ and $f_2$, we present the first of two experimental findings regarding CNOT chains in figure \ref{Plt: Paths Compared}, demonstrating differences in fidelities as a consequence of the four path orientations shown in figures \ref{Fig:Chain Path} and \ref{Fig:Chain Path Alts}.
	
	When comparing the plots in figure \ref{Plt: Paths Compared}, it is clear that there are some noticeable differences in fidelity rates between the various paths, some as large as 10 - 20$\%$ after length-10 chains.  Interestingly, a closer look at the data reveals distinct drops at certain chain lengths, particularly for paths $2$ (circles) and $3$ (diamonds), and $4$ (squares).  Upon further investigation into the location of each path's drop in fidelity, we find that they all occur at qubit 7 (see figure \ref{Fig:Chain Path}), which was later confirmed to have the lowest $T_1$ at the time of the experiments.  Thus, these results demonstrate how the location of a single noisy qubit can lead to varying algorithmic success rates based circuit configuration, which is a result that will be further demonstrated in later experiments.   Next, we now present CNOT chain results which showcase each of the circuit techniques in figure \ref{Fig:Chain Circuits} and their effectiveness at controlling the final state of the ancilla.

	\begin{figure}[h]                     
		\centering
		\includegraphics[scale=.55]{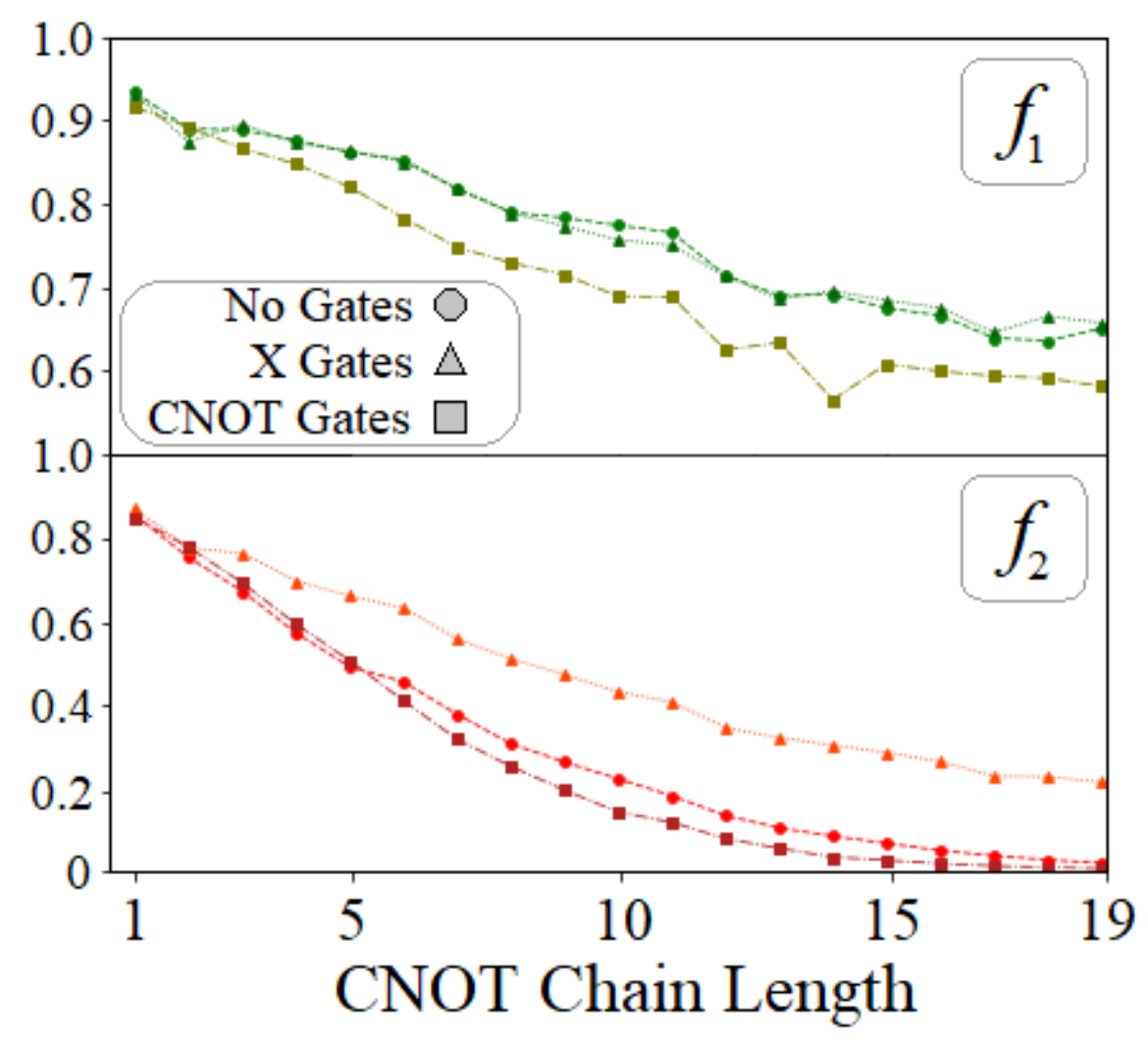}
		\caption{A comparison of fidelities for the three circuit types laid out in figure \ref{Fig:Chain Circuits}, averaged across all four path orientations.  The top plots demonstrate each circuit type's ability to successfully deliver the distant CNOT operation, while the bottom plots show how reliably each circuit handles the final state of the ancilla qubits.}
		\label{Plt: Chain Circuits Avg}
	\end{figure}
	
	The results shown in figure \ref{Plt: Chain Circuits Avg} represent average fidelities obtained from the four path orientations.  Beginning with $f_1$, the data shows that the presence of additional $X$ gates for resetting ancilla qubits has no impact when compared to using no gates, demonstrated by the overlap of the circle and triangle plots.  This agrees with what one would expect theoretically, given that chronologically the chain of CNOTs in both circuits are the same, therefore uninfluenced by any later gate operations on each ancilla qubit.  Additionally, the total lengths of time for each circuit are nearly identical, only differing by a single $X$ gate.  Conversely, because of the way in which the additional CNOT gates are staggered in the third circuit type (squares), we see a decrease in $f_1$ fidelity resulting from the fact that these extra gates ultimately delay the final measurement, effectively creating more time for decoherence errors on the control and target qubits.
	
	Now turning to the results for each circuit's $f_2$ rates, the bottom plot in figure \ref{Plt: Chain Circuits Avg} demonstrates each circuit type's ability to produce reliable ancilla states.  Using the case of no gates as a baseline, the data shows that the usage of $X$ gates can lead to a drastic improvement, while the use of CNOT gates has the opposite effect.  When comparing the plots for $X$ gates versus no gates, the data shows a widening gap as a function of chain length.   We can understand this trend as a result of $T_1$ decays, which become more problematic as the number of ancilla increases (more opportunities for collapse) as well as circuit length (more time that each qubit must maintain its excited state).  Conversely, the usage of additional $X$ gates immediately after each CNOT remedies this problem by effectively minimizing the time each ancilla qubit is subject to $T_1$ relaxation, regardless of chain length.  
	
	Lastly, the results for the circuit type utilizing CNOT gates shows the worst $f_2$ rates between the three.  Conceptually, resetting the ancilla qubits in this way is plagued with several issues, the worst of which being increased circuit length.  By requiring an entire second CNOT chain for resetting, ancilla qubits closer to the control must maintain their excited state for almost double that of the other two circuits.  Additionally, the success of each ancilla being properly reset is conditioned on the one prior, which means that a single intermediate $T_1$ relaxation is enough to disrupt the entire resetting process.  Although this circuit type has been shown to be the worst in both $f_1$ and $f_2$ fidelities, it is important to note that amongst the three circuit types it is the only one which can properly reset ancilla when the control qubit is in a superposition.  Thus,  despite the advantage in ancilla control provided from using $X$ gates, oftentimes the requirement of superposition in algorithms forces the use of additional CNOT gates.
	
	\section{Qubit Geometry \& Algorithm Design}
	
	Having now seen the extent to which a single CNOT operation can be reliably transmitted across ancilla qubits, the next question is how useful could such chains be for constructing larger circuits?  With limited connectivity on future NISQ devices being the expected standard, near term quantum circuits will need to critically rely on ancilla qubits and various techniques for algorithm implementation.  In this section we present several qubit geometries and circuit implementations for which we later present experimental results (see sections V and VI).
	
	\subsection{3 Qubit Geometry}
	
\begin{figure}[h]                     
		\centering
		\includegraphics[scale=.18]{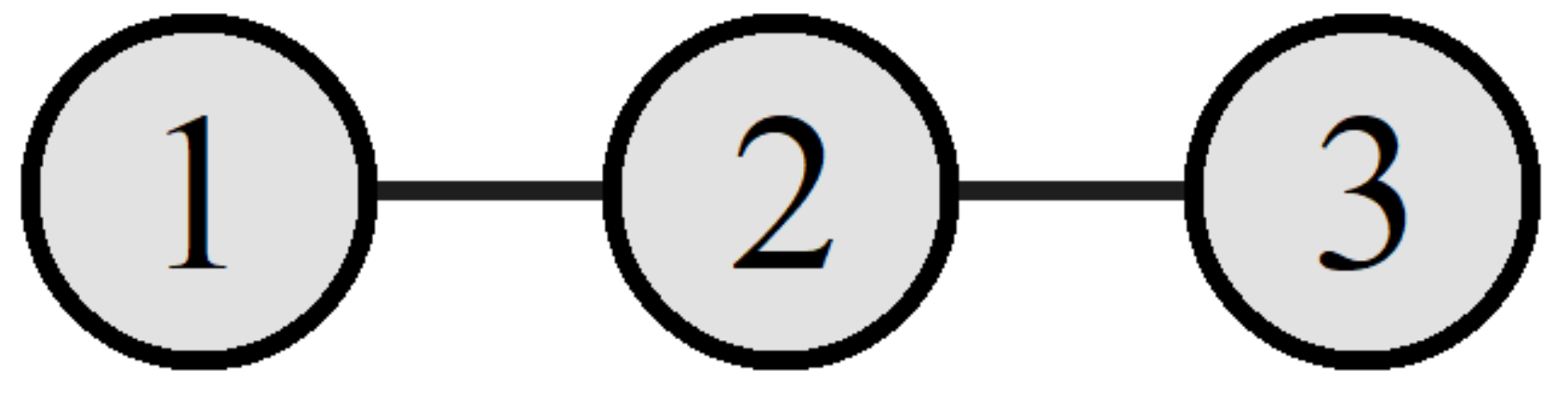}
		\caption{Three linearly connected qubits.}
		\label{Fig: 3 Geometry}
	\end{figure}

	Despite lacking any set of three directly interconnected qubits, the Poughkeepsie architecture possesses numerous combinations of three linearly connected qubits, as shown in figure \ref{Fig: 3 Geometry}.  Using qubits in this way to implement 3-qubit algorithms has the advantage of avoiding the need for any additional ancilla qubits, but becomes problematic when 2-qubit gate operations are required between the outer two qubits.  Compensating for this lacking connection requires additional 2-qubit gates through the central qubit, typically SWAP gates, which  consequently increase circuit length and noise susceptibility.
	
	When implementing 2-qubit gate operations between the two unconnected qubits in figure \ref{Fig: 3 Geometry}, the challenge lies in protecting the central qubit from additional noise while simultaneously ensuring that its quantum state is unaltered in the end.  The standard approach for implementing a 2-qubit gate between qubits $1$ and $3$ in figure \ref{Fig: 3 Geometry} would be to use SWAP gates, which effectively interchange the quantum state held on two qubits.  Through the use of SWAP gates, the quantum states held on distant qubits can be swapped onto qubits which possess a direct connection, allowing for the 2-qubit gate operation to occur.
	
	\begin{figure}[h]                     
		\centering
		\includegraphics[scale=.25]{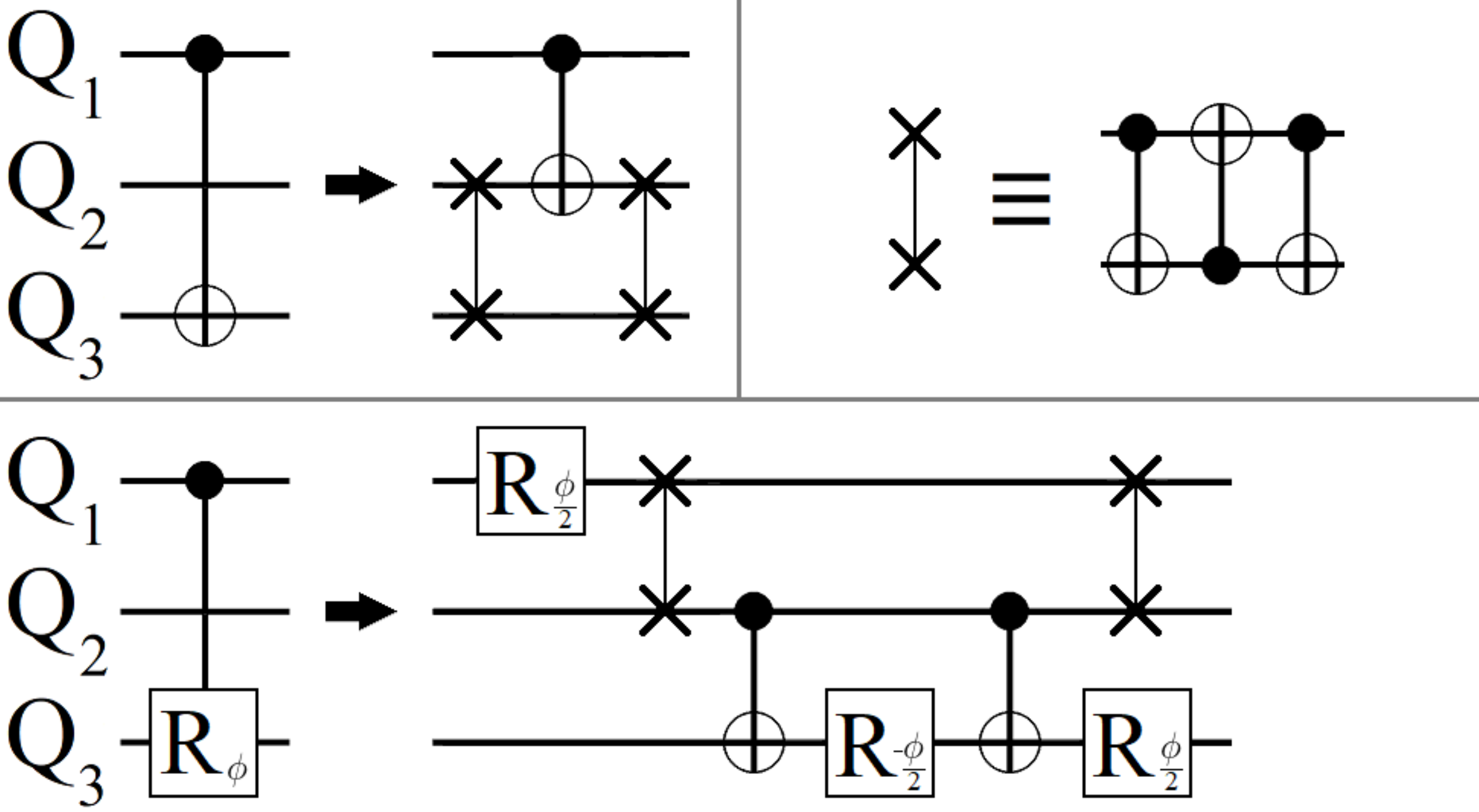}
		\caption{(top left) Circuit implementation for a CNOT gate between qubits $1$ and $3$, which are both connected to qubit $2$, but do not share a direct connection between themselves (see figure \ref{Fig: 3 Geometry}). (top right) Gate construction for a SWAP gate, consisting of three alternating CNOT gates. (bottom) Circuit implementation for a control-$R_{\phi}$ gate between qubits $1$ and $3$ for the same geometry.}
		\label{Fig: 3G 2Qubit Gates}
	\end{figure}
	
	The circuits shown in figure \ref{Fig: 3G 2Qubit Gates} are those tested in the coming experiments.  While not always optimal in circuit depth and gate count, each circuit is guaranteed to accomplish two things: 1) the successful implementation of the 2-qubit operation between the distant qubits, and 2) the central qubit is always returned back to its initial quantum state.  This second point comes at the cost of the second SWAP gate in each circuit, where notably this is where one typically looks to optimize when possible.

	\subsection{4 Qubit Geometry}
	
	If one wishes to avoid interchanging any computational qubits in order to compensate for missing connections, such as with the 3-qubit geometry, then one is forced to increase the size of the quantum system through the introduction of ancilla qubits.  Figure \ref{Fig: 4 Geometry} below illustrates one such solution, using a single central ancilla qubit to supply all 2-qubit gate operations between the surrounding three computational qubits.
	
	\begin{figure}[h]                     
		\centering
		\includegraphics[scale=.18]{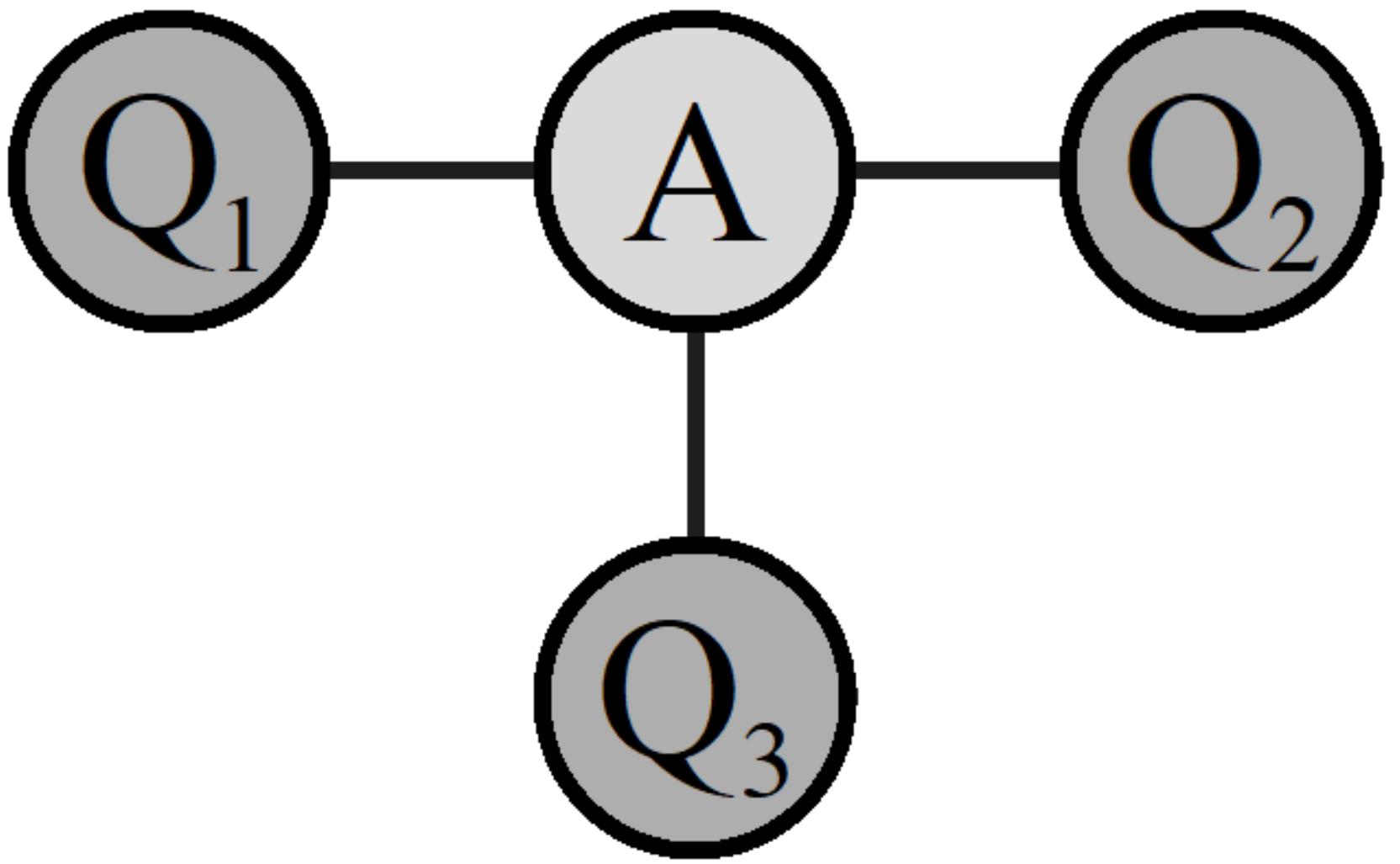}
		\caption{Four qubit geometry used for implementing 3-qubit algorithms throughout this study.  The central ancilla qubit allows for the implementation of 2-qubit gates between any two computational qubits, without interfering with the state of the third qubit.}
		\label{Fig: 4 Geometry}
	\end{figure}
	
	\begin{figure}[h]                     
		\centering
		\includegraphics[scale=.18]{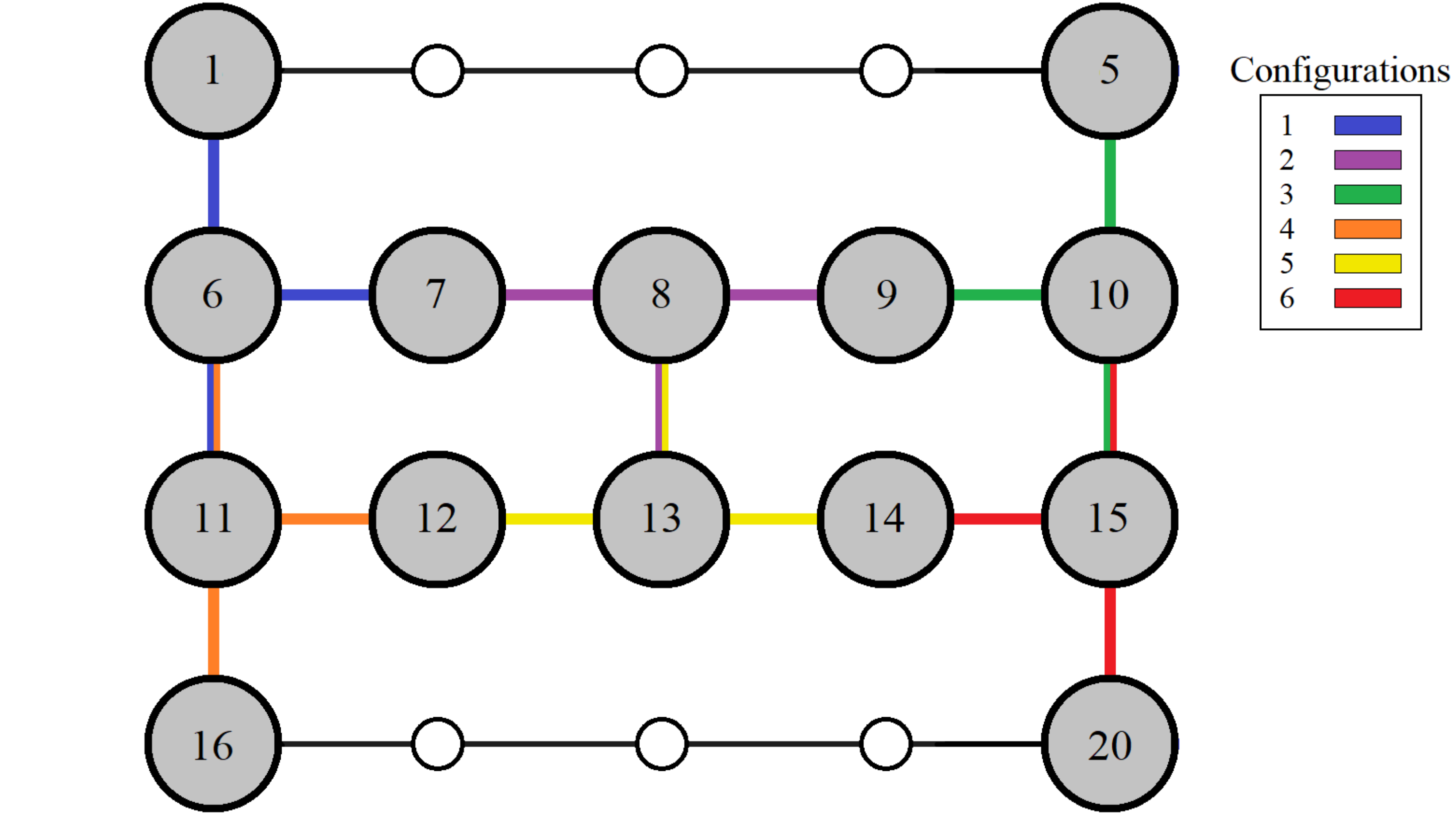}
		\caption{The six possible locations one can construct the 4-qubit configuration shown in figure \ref{Fig: 4 Geometry}}
		\label{Fig: 4 Geometry Locations}
	\end{figure}
	
	The 4-qubit geometry shown above represents an alternative to the problem of the 3-qubit geometry discussed earlier, using a single ancilla in order to avoid any additional SWAP gates on the computational qubits.  However, using a single ancilla for all 2-qubit gates requires that the state of the ancilla be properly reset after each usage in order to ensure the success of future gate operations.   Figure \ref{Fig: 4G 2Qubit Gate} below shows two examples for resetting the ancilla qubit when implementing a CNOT gate between computational qubits, where the choice for resetting depends on the presence of superposition on the control qubit. 
	
	\begin{figure}[h]                     
		\centering
		\includegraphics[scale=.25]{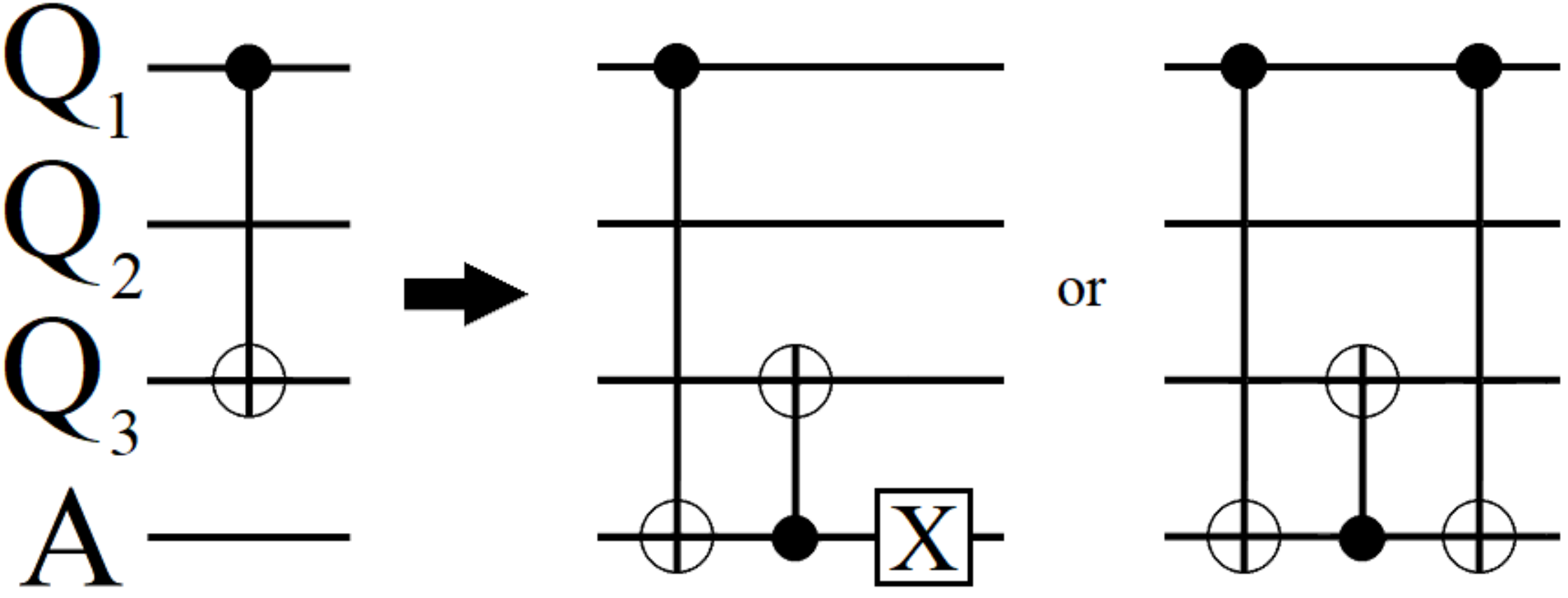}
		\caption{Example implementations of a CNOT gate between two computational qubits for the 4-qubit geometry (see figure \ref{Fig: 4 Geometry}).  Depending on whether or not the control computational qubit possesses superposition determines whether an $X$ or additional CNOT gate is necessary for resetting the state of the ancilla.}
		\label{Fig: 4G 2Qubit Gate}
	\end{figure}
	
	In comparing the 3 and  4-qubit geometries, the question becomes whether or not the additional ancilla qubit aids or hinders algorithmic success.  By requiring all 2-qubit gate operations traverse through a single qubit, a heavy burden is placed on the 4-qubit geometry's ancilla in terms of proper state manipulation and minimizing noise.  A similar burden is placed on the central qubit in the 3-qubit geometry, leading one to expect that a majority of algorithmic success is tied to these central qubits'  coherence properties and gate fidelities.
	
	\subsection{6 Qubit Geometries}
	
\begin{figure}[h]                     
		\centering
		\includegraphics[scale=.18]{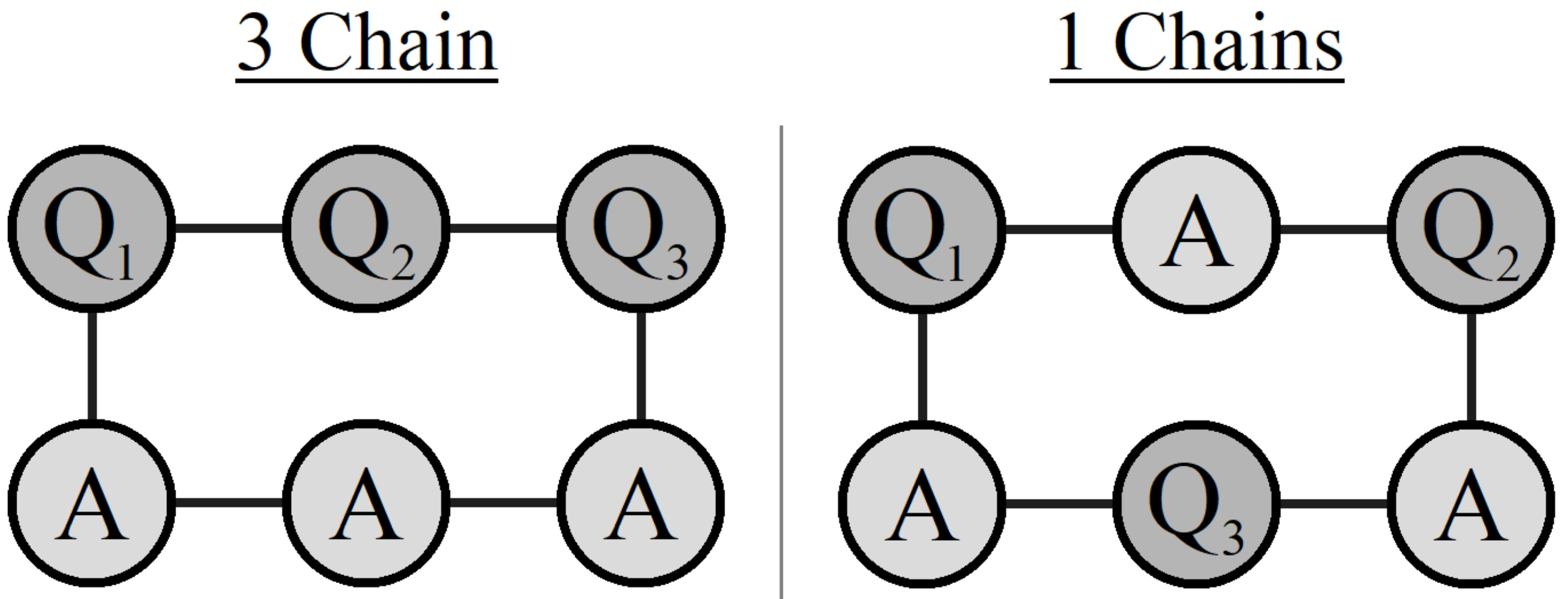}
		\caption{Six qubit geometries used for implementing 3-qubit algorithms throughout this study.  (left) Qubit geometry which uses three ancilla qubits to supplement the lack of connection between qubits Q$_1$ and Q$_3$. (right) Qubit geometry whereby each computational qubit is separated by a single ancilla.}
		\label{Fig: 6 Geometry}
	\end{figure}

	As the final qubit geometry size tested in this study, we present here two 6-qubit geometries, shown in figure \ref{Fig: 6 Geometry}, which are motivated by the respective strengths and weaknesses anticipated of the 3 and 4-qubit geometries already mentioned.   Specifically, the left circuit in figure \ref{Fig: 6 Geometry} (`3 Chain') contains the same connectivity between the three computational qubits as the 3-qubit geometry, sharing the advantage of having two-out-of-three direct connections.  Conversely, the right geometry (`1 Chains') uses one ancilla for all 2-qubit gate operations just as with the 4-qubit geometry, but avoids the issue of repeated single ancilla use by using a different ancilla between each computational qubit.  While not expected to outperform the other two configurations, the two geometries put forth in figure \ref{Fig: 6 Geometry} are designed to provide additional insight into the role of using ancilla in implementing algorithms over more sparse architectures.
	
	\begin{figure}[h]                     
		\centering
		\includegraphics[scale=.15]{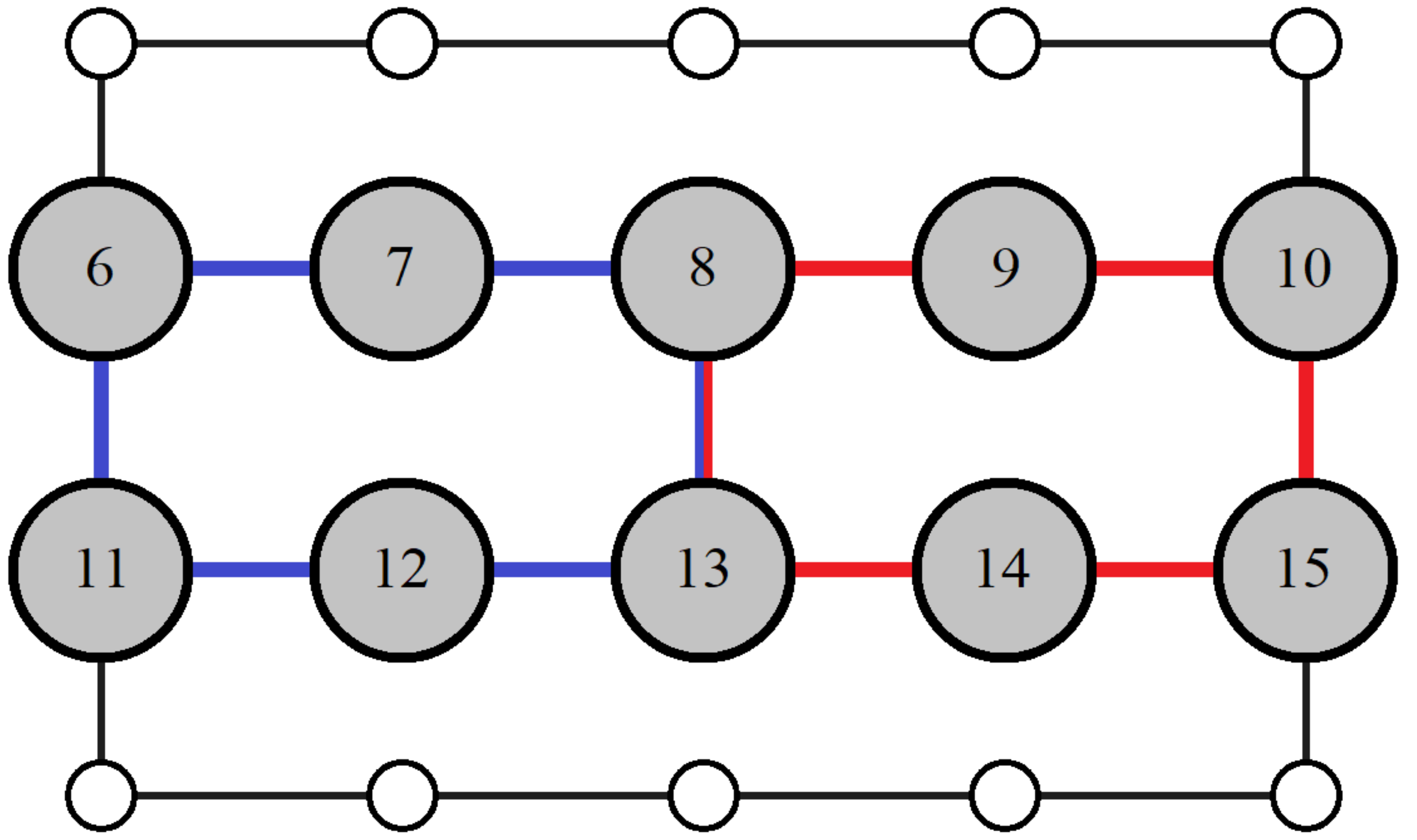}
		\caption{The two 6-qubit  geometries experimentally tested on the Poughkeepsie architecture.}
		\label{Fig: 6Q Loops}
	\end{figure}

	\section{CCNOT Experimental Results}
	
\begin{figure}[h]                     
		\centering
		\includegraphics[scale=.26]{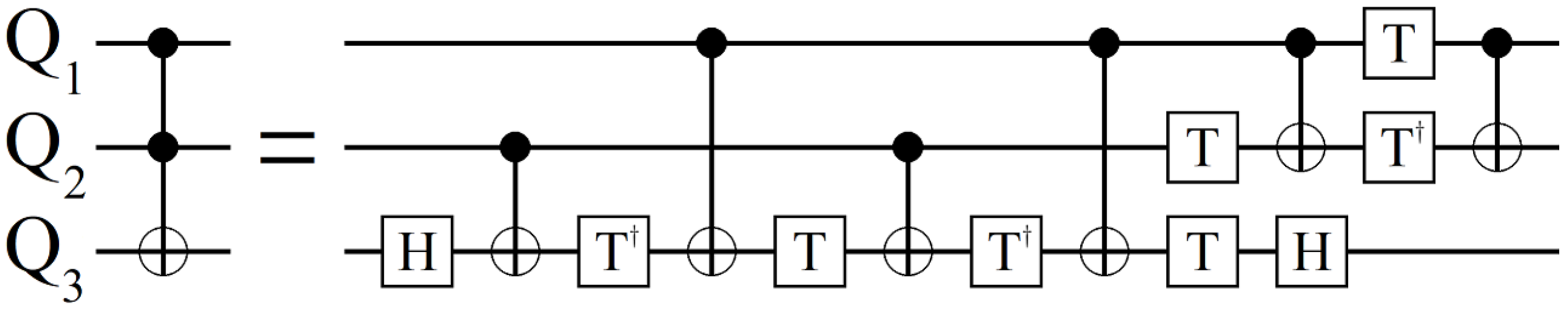}
		\caption{Quantum circuit for implementing a CCNOT gate operation \cite{toffoli2}.  $Q_1$ and $Q_2$ serve as the control qubits, while $Q_3$ is the target, receiving the equivalent of an X gate if both $Q_1$ and $Q_2$ are in the $|1\rangle$ state.}
		\label{Fig: CCNOT Circuit}
	\end{figure}

	For the progression of quantum computing, the importance of the CCNOT (Control-Control-$X$ gate) operation, also known as a Toffoli gate, cannot be understated as numerous quantum algorithms critically rely on its usage.  This includes algorithms which rely on oracles such as Grover's \cite{grover}, modular multiplication like Shor's \cite{shor2}, higher order control operators for Quantum Phase Estimation \cite{qpe}, or creating entanglement through mixing operators like in QAOA (Quantum Approximate Optimization Algorithm) \cite{qaoa} just to name a few.  Unlike previous operations studied up to this point however, the CCNOT gate is better thought of as a quantum circuit, consisting of several single and 2-qubit, shown in figure \ref{Fig: CCNOT Circuit}.
	
	In order to implement the quantum circuit outlined in figure \ref{Fig: CCNOT Circuit}, one necessary condition is that the three qubits be interconnected, requiring CNOT gates between the target and control qubits, and the two controls as well.  As already mentioned however, such connectivity does not exist on IBM's 20-qubit Poughkeepsie architecture, which means the circuit must be adapted to fit the various qubit geometries laid out in the previous section.
	
	\subsection{3 Qubit Results}
	
\begin{figure}[h]                     
		\centering
		\includegraphics[scale=.55]{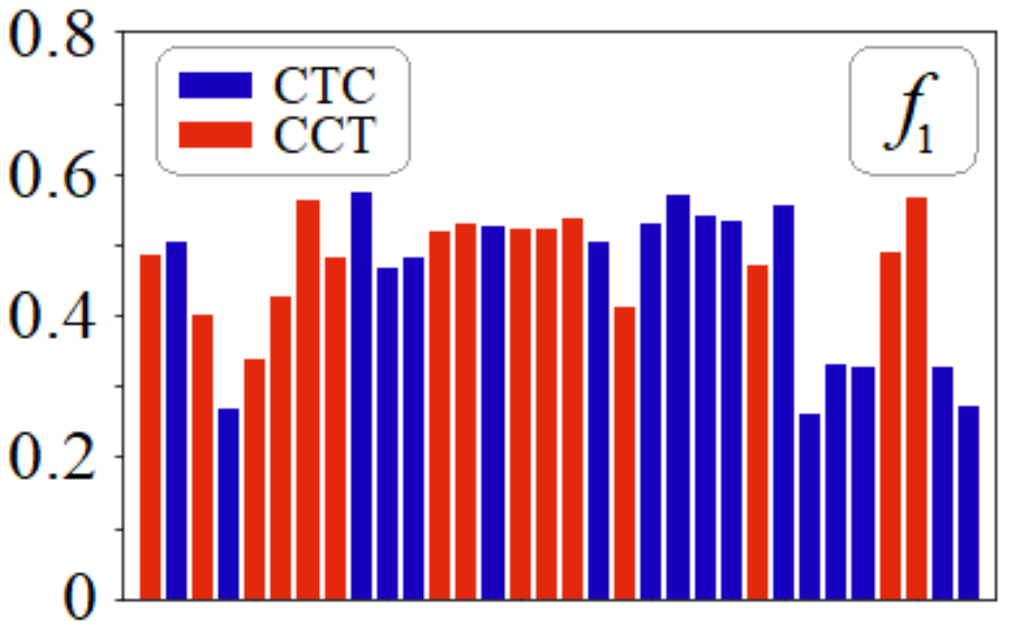}
		\caption{Highest fidelities found for the 32 possible 3-qubit combinations on the Poughkeepsie architecture, implementing the CCNOT circuit shown in figure \ref{Fig: CCNOT Circuit}.  The orientation which produced the higher fidelity is indicated by color, blue for the case where the central qubit was the target (CTC) and red for when it was an outer qubit (CCT).}
		\label{Fig: CCNOT 3 Qubits}
	\end{figure}

	In testing the CCNOT circuit using three linearly connected qubits, there are two unique configurations for which one can implement the control and target qubits.  Specifically, one can have the target qubit be either an outer (referred to as `CCT') or central (`CTC') qubit.  Due to the design of the CCNOT circuit, which requires exactly two CNOT gates between all three qubits, both configurations result in the same circuit depth and gate count when using SWAP gates to supplement the missing outer connection.  In total, the Poughkeepsie architecture possesses 32 possible 3-qubit combinations, all of which were tested using both configurations, and the results are shown in figure \ref{Fig: CCNOT 3 Qubits}.  In each trial, the control and target qubits are prepared in the $|1\rangle$ and $|0\rangle$ states respectively before passing through the CCNOT circuit.  Just as with the CNOT chain experiments, we are interested in the fidelity of the final measured state, whereby the target and both control qubits are all found to be in the $|1\rangle$ state.
	
	\begin{eqnarray}
	f_1 \hspace{.15cm} &\equiv&  \hspace{.15cm}  \big{|}\hspace{.05cm} \langle \hspace{.02cm}111\hspace{.02cm}|\hspace{.02cm}\textrm{Q}_1 \textrm{Q}_2 \textrm{Q}_3\hspace{.02cm}\rangle \hspace{.05cm} \big{|}^2
	\label{Eqn: CCNOT f1}
	\end{eqnarray}
	
	Figure \ref{Fig: CCNOT 3 Qubits} shows the fidelity rates found across the 96 tested CCNOT circuit implementations (three unique locations for the target qubit per each of the 32 total combinations).  As illustrated by the colors and numerical values, it is clear that no single combination of qubits or orientation is dominant in producing the best CCNOT fidelity.  While certain combinations produced worse fidelities as a result of noisier qubits, the data suggests that on average one can expect a successful CCNOT gate implementation on the order of 50-60\%, with only a select few noisy qubits reducing these values to around 25-40\%.
	
	\subsection{4 \& 6 Qubit Results}
	
	When analyzing the results for the 4 and 6-qubit geometry implementations of the CCNOT circuit, the addition of ancilla qubits requires the tracking of the $f_2$ metric in addition to $f_1$.  The interest is once again in how well each qubit configuration can reliably reset their respective ancilla qubits back to the $|0\rangle$ state through the use of either $X$ or CNOT gates.
	
	\begin{eqnarray}
	\textrm{4-Qubit:} \hspace{.3cm} f_2 \hspace{.1cm} &\equiv& \hspace{.1cm}  \big{|}\hspace{.05cm}   \langle \hspace{.02cm}111\hspace{.02cm}|\hspace{.02cm}\textrm{Q}_1 \textrm{Q}_2 \textrm{Q}_3\hspace{.02cm}\rangle \otimes \langle \hspace{.02cm}0\hspace{.02cm}|\hspace{.02cm}\textrm{A}\hspace{.02cm}\rangle  \hspace{.05cm} \big{|}^2  \\
	\textrm{6-Qubit:} \hspace{.3cm} f_2 \hspace{.1cm} &\equiv& \hspace{.1cm}  \big{|}\hspace{.05cm}   \langle \hspace{.02cm}111\hspace{.02cm}|\hspace{.02cm}\textrm{Q}_1 \textrm{Q}_2 \textrm{Q}_3\hspace{.02cm}\rangle \otimes \langle \hspace{.02cm}000\hspace{.02cm}|\hspace{.02cm}\textrm{A}_1 \textrm{A}_2 \textrm{A}_3\hspace{.02cm}\rangle  \hspace{.05cm} \big{|}^2 \hspace{.5cm}
	\label{Eqn: CCNOT f2 }
	\end{eqnarray}
	
	Beginning with the 4-qubit geometry, two cases for handling the resetting of the ancilla qubit were tested, corresponding to the two implementations shown in figure \ref{Fig: 4G 2Qubit Gate}.  For each implementation, all six possible 4-qubit combinations on the Poughkeepsie architecture (see figure \ref{Fig: 4 Geometry Locations}) were experimentally tested, setting each of the three outer qubits as the target.  The average results for each qubit combination are shown below in figure \ref{Fig: CCNOT 4 Qubits}.
	
	\begin{figure}[h]                     
		\centering
		\includegraphics[scale=.52]{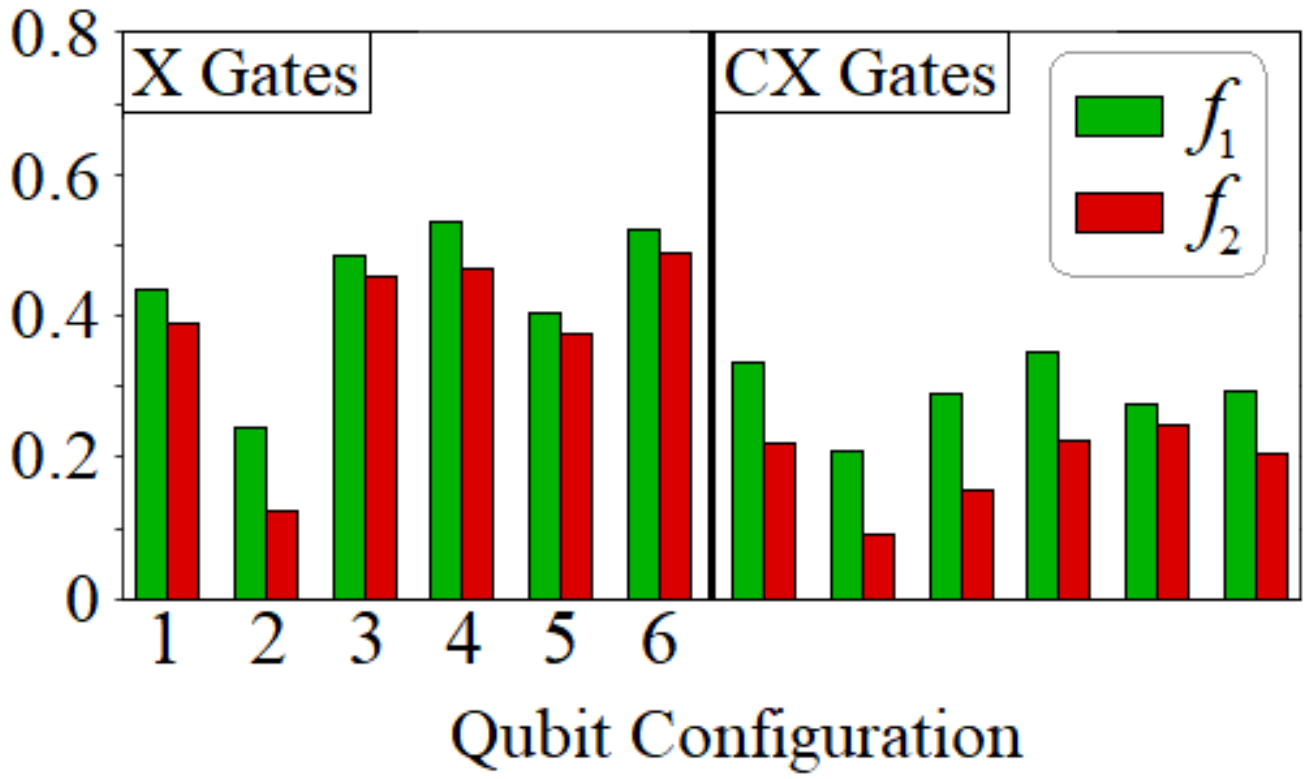}
		\caption{$f_1$ (green) and $f_2$ (red) rates for the 4-qubit implementations of the CCNOT gate. The two sets of data correspond to using $X$ gates (left) for resetting the ancilla qubit, versus using additional CNOT gates (right).}
		\label{Fig: CCNOT 4 Qubits}
	\end{figure}
	
	When looking at the results in figure \ref{Fig: CCNOT 4 Qubits}, it is clear that the use of $X$ gates for resetting the central ancilla qubit produce noticeably higher fidelities, both for $f_1$ and $f_2$.  This is in agreement with the CNOT chain results from earlier (figure \ref{Plt: Chain Circuits Avg}), once again highlighting the cost in ancilla control when forced to implement CNOT gates to account for superposition.  If we now compare these results to those of figure \ref{Fig: CCNOT 3 Qubits}, we find that the fidelities between the 3 and 4-qubit geometries using $X$ gates are comparable, with a slight edge going to the 3-qubit geometry.  The closeness in the two results suggests that the use of an ancilla qubit, versus SWAP gates through a computational qubit, is a viable approach for CCNOT algorithm design.  However, this viability is lost when superpositions must be accounted for, which are handled automatically by SWAP gates for the 3-qubit geometry, but require CNOT gates for the 4-qubit geometry.
	
	Proceeding now to the 6-qubit geometries, figure \ref{Fig: CCNOT 6 Qubits} below shows the full results for the two implementations illustrated in figure \ref{Fig: 6 Geometry}, once again separated into the metrics $f_1$ and $f_2$.  Overall, the data shows that the `3 Chain' circuit implementation yields comparable fidelity rates to the 4-qubit configurations using $X$ gates.  However, the use of a single ancilla between each computational qubit shows a dramatic decrease.
	
	\begin{figure}[h]                     
		\centering
		\includegraphics[scale=.45]{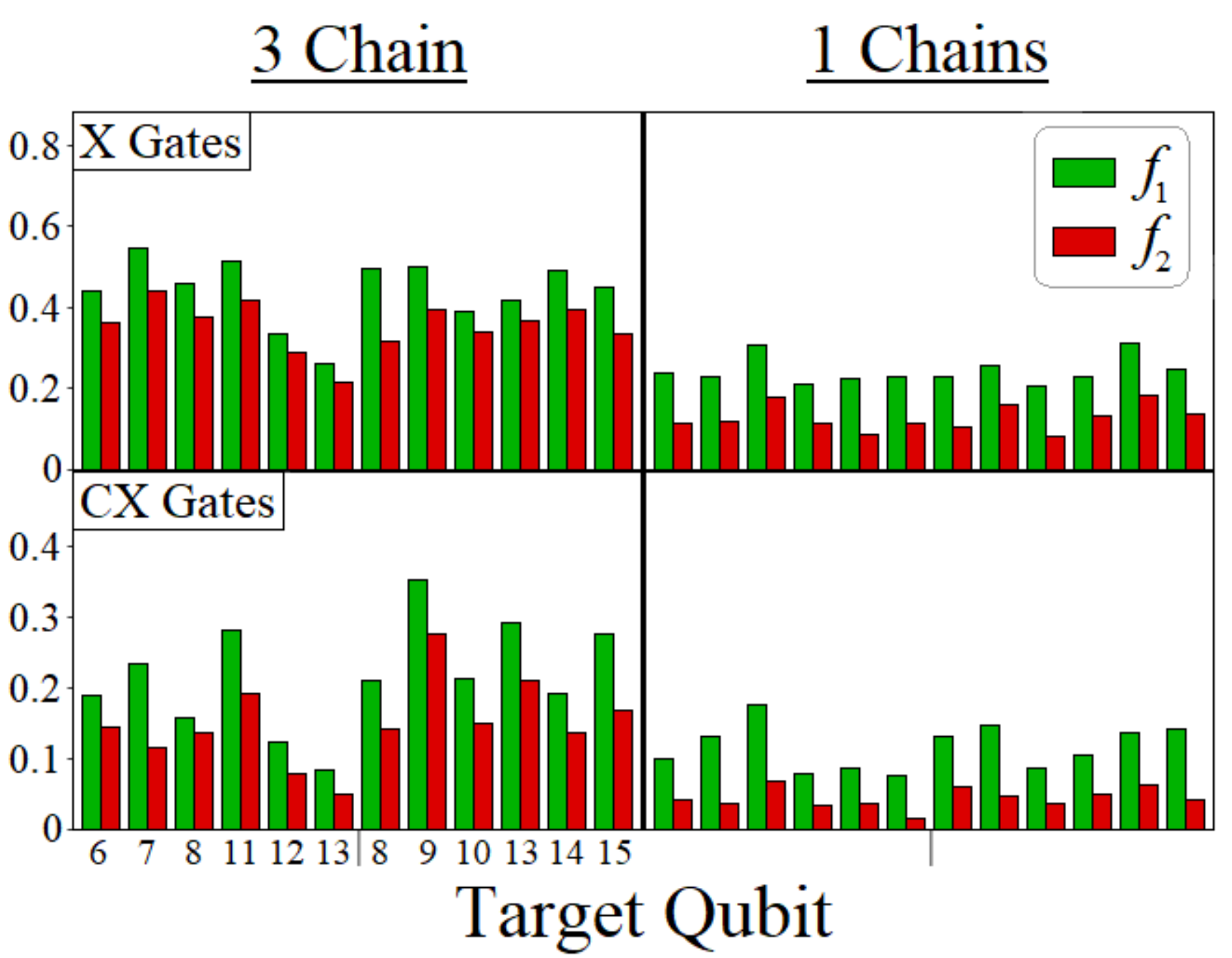}
		\caption{Fidelities $f_1$ and $f_2$ for the `3 Chain' (left) and `1 Chains' (right) circuits.  Each bar represents one of the twelve possible qubit configurations, denoted by the qubit in each 6-qubit ring acting as the target qubit for the CCNOT action (see figures \ref{Fig: 6 Geometry} and \ref{Fig: 6Q Loops}).}
		\label{Fig: CCNOT 6 Qubits}
	\end{figure}
	
	When comparing the two circuit designs, which only differ in the manner in which the ancilla qubits are distributed, it is important to note that both circuit implementations use the same total number of CNOT and $X$ gates, producing near identical circuit depths.  Additionally, in both cases each ancilla qubit is called upon exactly twice, followed immediately by $X$ or CNOT gates for resetting back to the $|0\rangle$ state.  These consistencies suggest that the results shown in figure \ref{Fig: CCNOT 6 Qubits} are then attributable to the $\textit{way}$ in which the CNOT gates are distributed throughout the circuits.  
	
	More specifically, if we return to the results of the CNOT Chain experiment, and focus on the fidelities found for chains of $1$ and $3$ ancilla qubits, we can approximate the average $f_1$ fidelities to be $0.9$ and $0.85$ respectively.  If we now compare the way in which each circuit requires CNOT chains in order to supplement missing connections, we find that the `3 Chain' configuration only needs two successful chains, while the `1 Chains' configuration is reliant on six.  Using our example approximate fidelities, this means that we would expect the probability of success from each circuit to be ($0.9$)$^6$ and ($0.85$)$^2$ respectively, heavily favoring the `3 Chains' circuit design.  Thus, in determining how best to arrange computational and ancilla qubits for algorithm design, the results shown above suggest that grouping computational qubits closer together, in favor of fewer but longer ancilla chains, will lead to better algorithmic success.
	
	\section{Quantum Fourier Transformation}
	
	Having just seen the varying degrees to which the Poughkeepsie architecture can handle CCNOT circuits, we now turn to another critical subroutine for quantum computing, the Quantum Fourier Transformation (QFT).  In this section, we present experimental results which demonstrate the reliability with which one can successfully perform a 3-qubit QFT using the various qubit geometries outlined earlier.  At its core, the QFT is the quantum equivalent to the Discrete Fourier Transformation, applied to a quantum state.  The power of the QFT lies in its ability to apply up to $2^{N}$ unique phases across the various components of a quantum state, where $N$ is the number of qubits.  The core element necessary to any successful QFT is the control-$R_{\phi}$ gate, which applies an arbitrary phase to a target qubit's $|1\rangle$ state, conditional on a control qubit.
	
	\begin{eqnarray}
	R_{\phi} \hspace{.02cm} |1\rangle \otimes ( \hspace{.04cm} \alpha \hspace{.02cm}|0\rangle + \beta \hspace{.02cm} |1\rangle \hspace{.04cm} ) \hspace{.1cm} = \hspace{.1cm}  |1\rangle \otimes ( \hspace{.04cm} \alpha \hspace{.02cm}|0\rangle + e^{i\phi} \hspace{.02cm}\beta \hspace{.02cm}|1\rangle \hspace{.04cm} )
	\label{Eqn: Control Phase Gate}
	\end{eqnarray}
	
	Just like the CCNOT circuit, the QFT requires full connectivity between all qubits.  The standard quantum circuit for a 3-qubit QFT is shown below in figure \ref{Fig: QFT Circuit} (technically a QFT$^{\dagger}$ circuit, which we discuss in the coming section), which we adapt accordingly for the various qubit geometries.  When physically implementing these 3-qubit QFT's, note that the true gate count for each control-$R_{\phi}$ gate includes additional CNOT and $R_{\phi}$ gates.  As a result, the circuit depth and total gate count for the 3-qubit QFT turns out to be comparable to that of the CCNOT circuit, which in turn will provide some insight when comparing the success rates between the two.
	
	\begin{figure}[h]                     
		\centering
		\includegraphics[scale=.3]{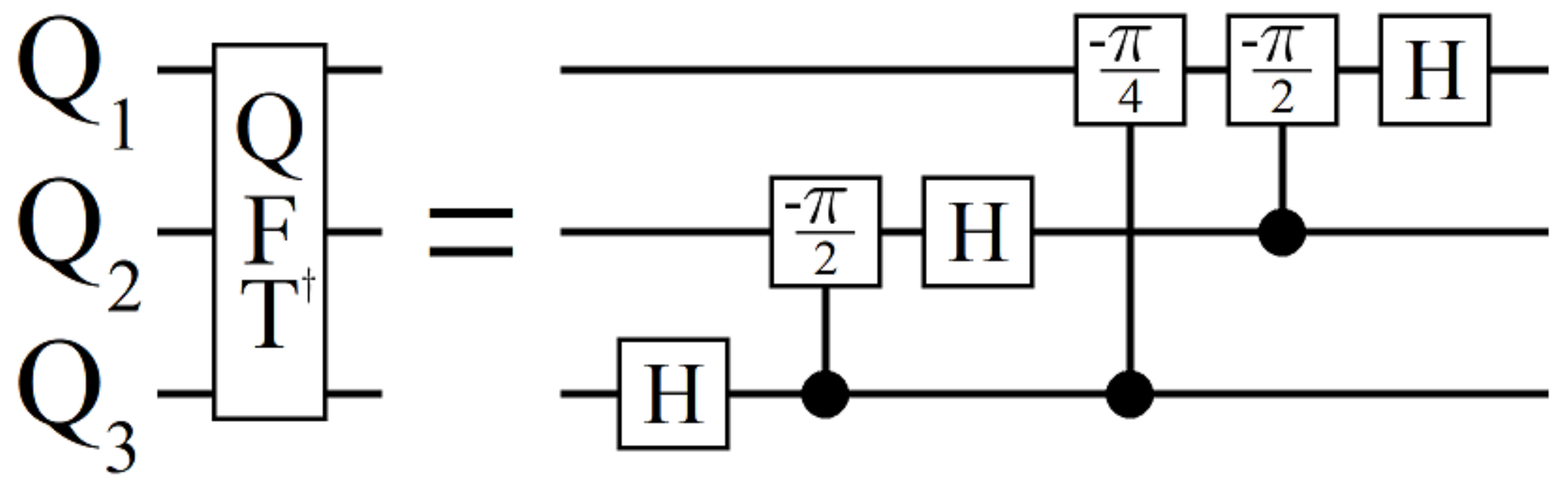}
		\caption{QFT$^{\dagger}$ circuit for three qubits.  The QFT$^{\dagger}$ shown above is the circuit tested on the IBM 20-qubit architecture, identical to the QFT circuit in both total gate count and circuit depth, differing only in gate order and phase values.}
		\label{Fig: QFT Circuit}
	\end{figure}
	
	\subsection{Testing QFT$^{\dagger}$ With Phase Estimation}
	
	In order to isolate and benchmark the success of the QFT$^{\dagger}$ circuit in figure \ref{Fig: QFT Circuit} in a way similar to the previous sections, one ideally needs an experiment whereby the effect of the QFT$^{\dagger}$ produces a single desirable final state.  However, unlike the CCNOT operation whose effect is directly observable by means of the target qubit, the QFT$^{\dagger}$ is a more versatile quantum operation whose effect ranges widely based on the state of the qubits it is applied to.  To this end, we quantify the fidelity of our QFT$^{\dagger}$ implementations in a manner analogous to the Quantum Phase Estimation Algorithm (QPE) \cite{qpe,book2}, whereby the effect of the final QFT$^{\dagger}$ leaves all of the qubits in a final state containing no superposition.  By creating very particular superposition states just prior to the QFT$^{\dagger}$ operation, we are guaranteed to have theoretical $|0\rangle$ and $|1\rangle$ final states for the computational qubits, with which we can then use to compute fidelities $f_1$ and $f_2$.
	
\begin{figure}[h]                     
		\centering
		\includegraphics[scale=.3]{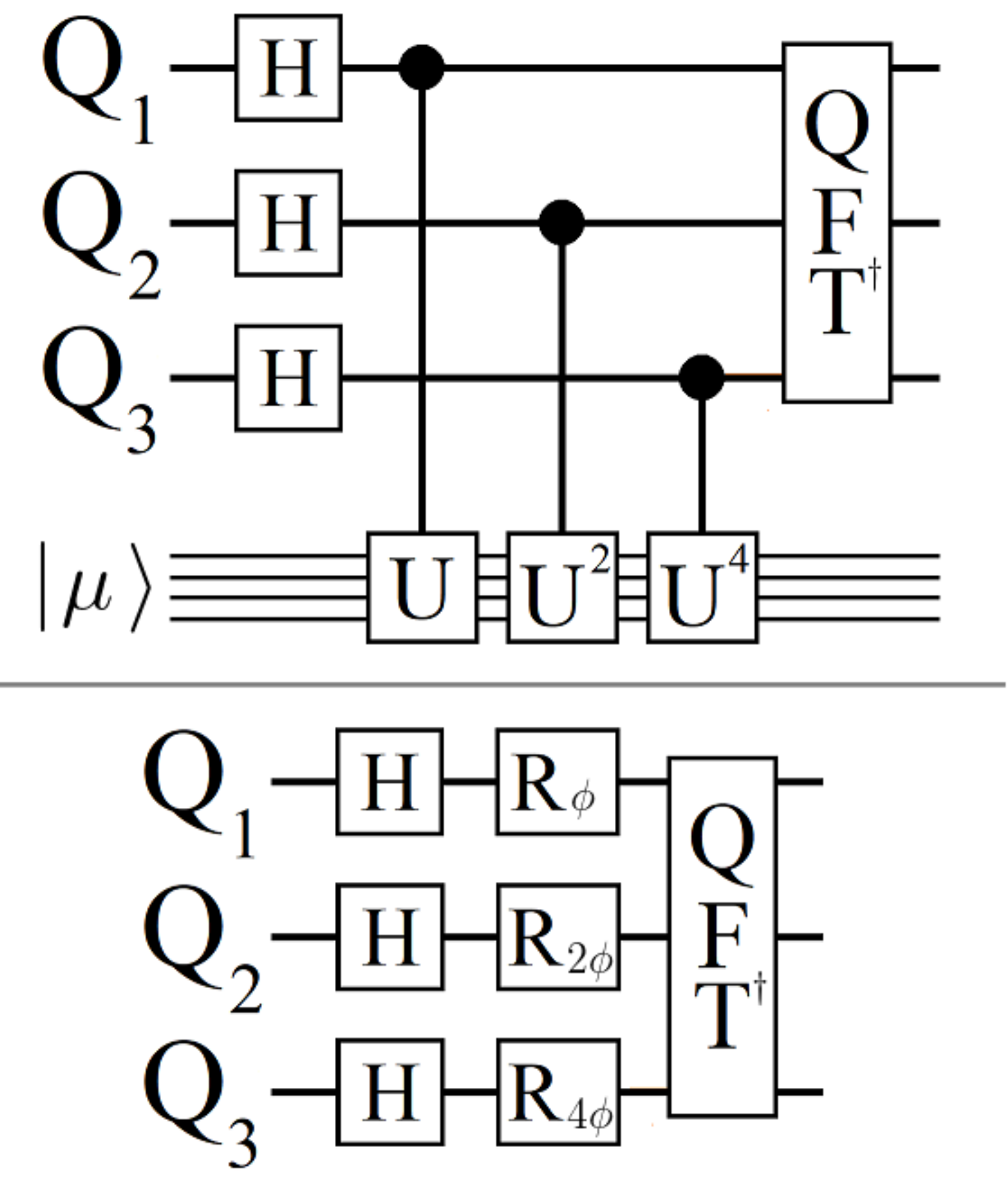}
		\caption{(top) Quantum circuit for a 3-qubit Quantum Phase Estimation Algorithm. (bottom) A circuit which mimics the effect of the control-U operations through the use of single qubit rotation gates.}
		\label{Fig: QPE Circuit}
	\end{figure}

	Quantum Phase Estimation is a quantum algorithm which uses a control-U operation, along with one of its eigenstates $|\mu \rangle$, in order to detect some unknown eigenphase $e^{i \theta}$.  An example QPE is shown below in the top circuit of figure \ref{Fig: QPE Circuit}.  Creating such a circuit is typically very challenging, as both the implementation of arbitrary control-U operators and their eigenstates require clever circuit design.  For the purpose of our QFT benchmarking however, we apply the core idea of the QPE in a much simpler form, effectively achieving the states resulting from the control-U operations acting on $|\mu \rangle$ with only $R_{\phi}$ gates, illustrated by the bottom circuit in figure \ref{Fig: QPE Circuit}.
	
	By using single qubit rotation gates to initialize the computational qubits, we are able to prepare quantum states just prior to the QFT$^{\dagger}$ with high fidelities, minimizing any additional noise not caused by the QFT$^{\dagger}$ circuit.  Additionally, the use of $R_{\phi}$ gates allows for the creation of a wider range of states than typically achievable through the use of control-U gates, which we in turn use for further insight in the viability of the QFT$^{\dagger}$ circuit in later experiments.
	
	\subsection{Perfect Phase Detection}
	
	For the case of a 3-qubit QFT$^{\dagger}$, there are exactly eight choices for $\phi$ such that the bottom circuit in figure \ref{Fig: QPE Circuit} will result in a final state containing no superposition.  These eight values of $\phi$ span an even distribution from $0$ to $\frac{7\pi}{4}$, corresponding to the eight unique quantum states from $|000\rangle$ to $|111\rangle$.  These eight states will serve as the desired final measurements for determining fidelities:
	
	\begin{eqnarray}
	f_1  &\equiv&    \big{|}\hspace{.05cm} \langle \hspace{.02cm}\textrm{Q}_1' \textrm{Q}_2' \textrm{Q}_3'\hspace{.02cm}|\hspace{.02cm}\textrm{Q}_1 \textrm{Q}_2 \textrm{Q}_3\hspace{.02cm}\rangle \hspace{.05cm} \big{|}^2  \\
	\textrm{4-Q:} \hspace{.1cm} f_2  &\equiv&   \big{|}\hspace{.05cm}   \langle \hspace{.02cm}\textrm{Q}_1' \textrm{Q}_2' \textrm{Q}_3'\hspace{.02cm}|\hspace{.02cm}\textrm{Q}_1 \textrm{Q}_2 \textrm{Q}_3\hspace{.02cm}\rangle \otimes \langle \hspace{.02cm}0\hspace{.02cm}|\hspace{.02cm}\textrm{A}\hspace{.02cm}\rangle  \hspace{.05cm} \big{|}^2   \\
	\textrm{6-Q:} \hspace{.1cm} f_2 &\equiv&   \big{|}\hspace{.05cm}   \langle \hspace{.02cm}\textrm{Q}_1' \textrm{Q}_2' \textrm{Q}_3'\hspace{.02cm}|\hspace{.02cm}\textrm{Q}_1 \textrm{Q}_2 \textrm{Q}_3\hspace{.02cm}\rangle \otimes \langle \hspace{.02cm}000\hspace{.02cm}|\hspace{.02cm}\textrm{A}_1 \textrm{A}_2 \textrm{A}_3\hspace{.02cm}\rangle  \hspace{.05cm} \big{|}^2  \hspace{.7cm}
	\label{Eqn: QFT Fidelities} 
	\end{eqnarray}
	
	In the QFT$^{\dagger }$ fidelity results to come, the highest fidelity rates from experiments \ref{Fig: CCNOT 3 Qubits} - \ref{Fig: CCNOT 6 Qubits} were used to determine which qubits to experimentally test.  Specifically, the top three qubit combinations for each geometry which yielded the highest fidelities were tested and then averaged together.  For the 3-qubit geometries, the top three qubit combinations for both control-target orientations were tested.  For the 6-qubit geometries, only the `3 Chain' orientation was tested (preliminary results showed once again a significant decrease in fidelity rates for the `1 Chains' orientation).  And finally, because the QFT$^{\dagger}$ circuit is always acting on a superposition state of the computational qubits, both the 4 and 6-qubit geometries require CNOT gates for resetting ancilla qubits.
	
	\begin{figure}[h]                     
		\centering
		\includegraphics[scale=.5]{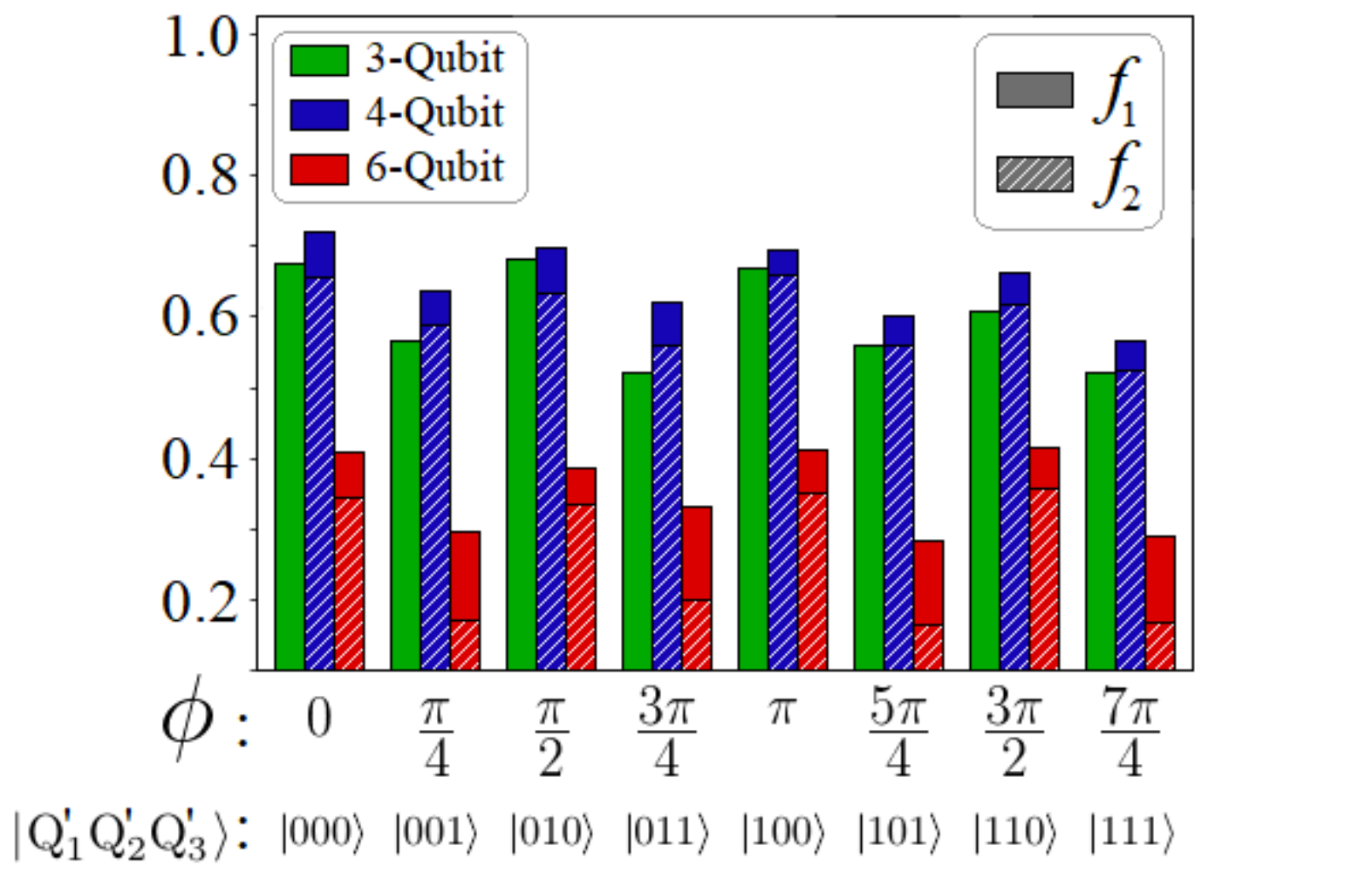}
		\caption{(solid fill) $f_1$ fidelities found for each qubit geometry, demonstrating each geometry's ability to produce and measure the eight desired final states resulting from the QPE circuit (bottom circuit in figure \ref{Fig: QPE Circuit}). For the 4 and 6-qubit results, $f_2$ (dashed fill) fidelity rates are also shown, highlighting each respective geometry's ability to reliably reset ancilla qubits. }
		\label{Fig: QFT Perfects}
	\end{figure}
	
	Beginning with $f_1$, the results shown in figure \ref{Fig: QFT Perfects} reveal that the 4-qubit geometry lead to the overall highest QFT$^{\dagger}$ fidelities across all eight phases.  In addition to the high $f_1$ rates, the accompanying high $f_2$ values suggests that one could also reliably perform further gate operations after the QFT$^{\dagger}$.  Behind the 4-qubit geometry we find that the 3 and 6-qubit geometries produced fidelities of the order 55-65\% and 30-40\% respectively.  
	
	Based on the results from the CCNOT experiments, the higher fidelity rates of the 4-qubit geometry may come as a surprise at first glance.  However, in analyzing the two top performing geometries and their circuit implementations, the key to the 4-qubit geometry's success lies in the ordering of the control phase gates.  Specifically, the $-\frac{\pi}{2}$ and $-\frac{\pi}{4}$ control-$R_{\phi}$ gates which happen in succession, originating from the same computational qubit, allow for a slight optimization in the 4-qubit circuit. 
	
	\begin{figure}[h]                     
		\centering
		\includegraphics[scale=.25]{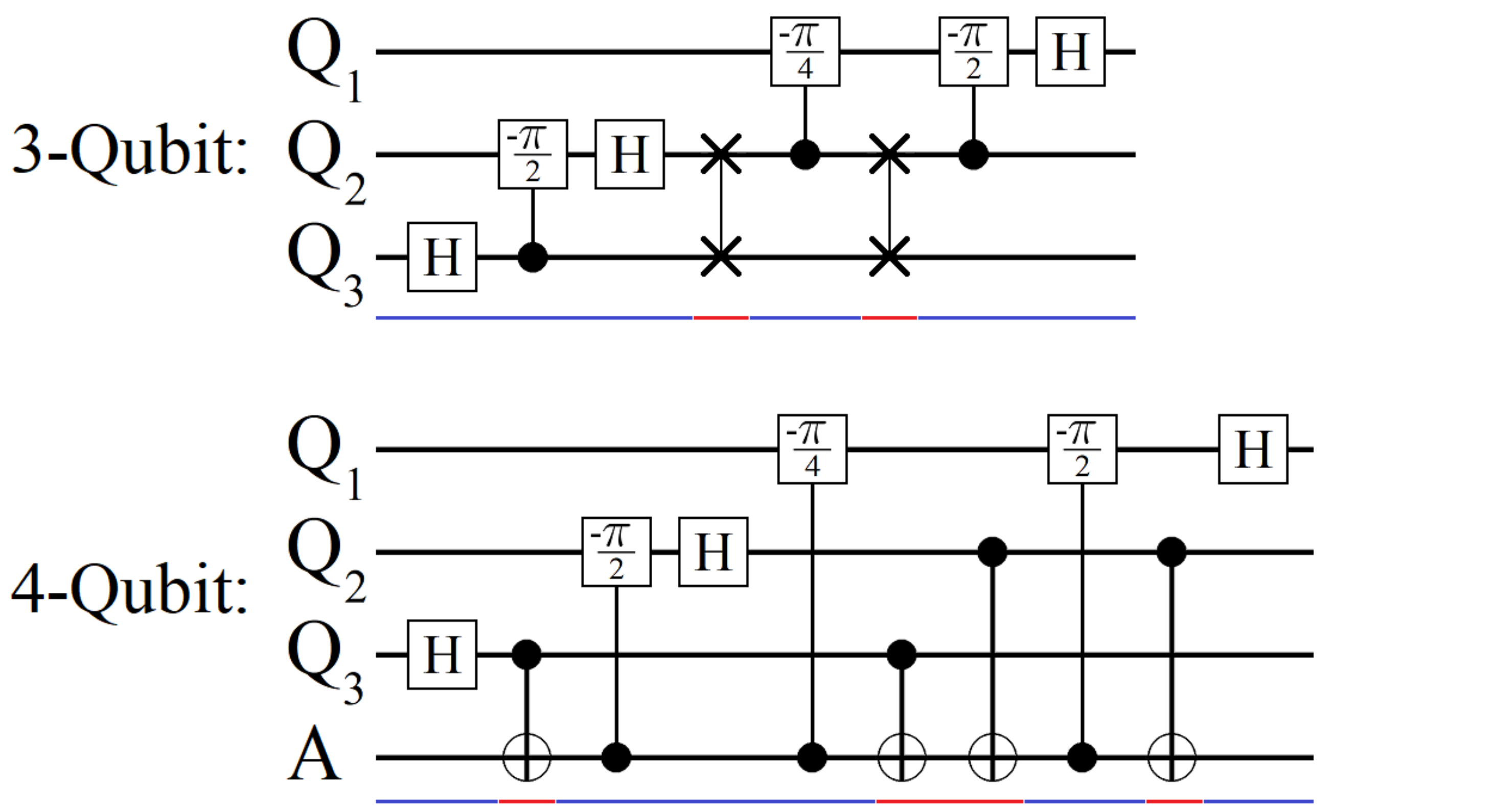}
		\caption{Circuit implementations of a 3-qubit QFT$^{\dagger}$, subject to the connectivity restraints outlined in figures \ref{Fig: 3 Geometry} and \ref{Fig: 4 Geometry}.  The blue and red underscores to each circuit highlight the gates which are the same (blue) and different (red) between the two circuits.  In total, the 4-qubit geometry achieves the QFT$^{\dagger}$ operation with two fewer CNOT gates.}
		\label{Fig: QFT 3 vs 4}
	\end{figure}
	
	As illustrated in figure \ref{Fig: QFT 3 vs 4}, the difference in implementation between the 3 and 4-qubit geometries boils down to the extra CNOT gates necessary to to compensate for each configuration's lacking connectivity. While the 3-qubit geometry requires two SWAP gates, for a combined total of six additional CNOT gates, the 4-qubit geometry only requires four.  Typically each control gate would require two CNOTs for resetting, but since two of them originate from the same computational qubit with no gates in between, there is no need to reset the ancilla qubit back down to the $|0\rangle$ state after the first $-\frac{\pi}{2}$ gate.  Thus, the results of figure \ref{Fig: QFT 3 vs 4} demonstrate that the use of an ancilla qubit can potentially be used to optimize circuit depth and consequently improve algorithm success.
	
	In addition to the qubit geometry discrepancies, a second interesting result emerging from the data reveals an alternating pattern in fidelities between phases.  This pattern is present across all three qubit geometries, suggesting that this phenomenon is inherently linked to QPE itself.   One possible explanation for this trend could be in the complexity of the quantum state just prior to the QFT$^{\dagger}$ circuit.  Specifically, the superposition states created from the even integer cases of $\frac{\pi}{4}$ contain at most four unique phases: $\frac{1}{8}$, $-\frac{1}{8}$, $\frac{i}{8}$, and $\frac{-i}{8}$ across the eight computational basis states.  Conversely, the odd integer cases contain four additional phases: $\frac{\pm 1 \pm i}{4}$, producing superposition states where each basis state has a unique phase.  Since these quantum states have more relative phases between the eight computational basis states, it is possible that they are more sensitive to noise and errors, leading to lower fidelity rates after the QFT$^{\dagger}$ circuit.

	\subsection{Continuous Phase Detection}
	
\begin{figure}[h]                     
		\centering
		\includegraphics[scale=.5]{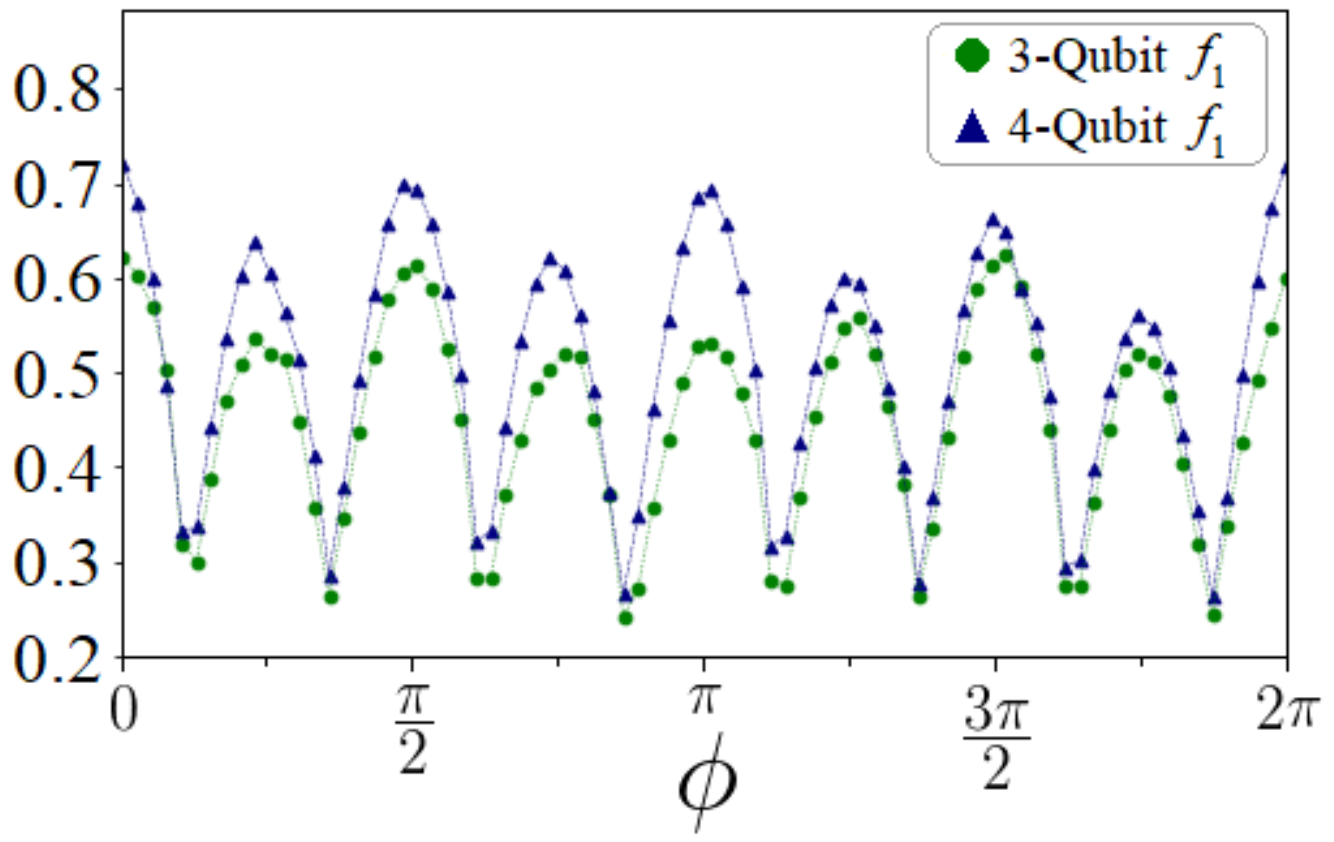}
		\caption{ QPE fidelity rates for the 3 (green circles) and 4-qubit (blue triangles) geometries as a function of phase $\phi$.  Each data point represents the measured percentage of states corresponding to the nearest `perfect phase' value for $\phi$ (see figure \ref{Fig: QFT Perfects}).}
		\label{Fig: QFT Continuous}
	\end{figure}

	Following from the data trends revealed in the previous section, we now present experimental results which are motivated by a more realistic usage of the QPE Algorithm.  Specifically, we present results which extend the data shown in figure \ref{Fig: QFT Perfects}, testing for intermediate values of $\phi$, ultimately attempting to detect phases which do not match up perfectly with one of the $2^N$ basis states created from the number of qubits being used.  Detecting these `non-perfect' phases comes with an inherent probability of failure, even for a noiseless quantum computer, as the resulting final states from the QFT$^{\dagger}$ now contain superposition.  Consequently, one expects lower fidelities in regions of $\phi$ between the $2^N$ perfect phases, with the lowest points being exactly halfway between each perfect phase (approximately 40\% for a noiseless 3-qubit QPE).  Figure \ref{Fig: QFT Continuous} confirms this trend, illustrating fidelity swings of nearly 50\% for changes in $\phi$ as little as $\frac{\pi}{16}$. 
	
	The $f_1$ rates illustrated in figure \ref{Fig: QFT Continuous} are in agreement with those found in the previous experiment, showing fidelities on the order of 60 to 75\% around the eight perfect phases.  Additionally, we once again see the alternating peaks in success between the odd and even integers of $\frac{\pi}{4}$, now extended to the nearby regions of $\phi$ as well. 
	
	When comparing the data from figure \ref{Fig: QFT Continuous} to what one would expect from a noiseless quantum computer, the quantity of interest here is the way in which noise affects the full range of phase values.  Specifically, since the anticipated results have less than $100\%$ theoretical fidelities, one might anticipate the manner in which noise impacts these fidelities in one of two ways.  Supposing one finds a $75\%$ fidelity for the perfect phases, would the presence of noise cause intermediate values of $\phi$ to similarly yield $75\%$ of their theoretical maximums, or simply result in a flat $25\%$ reduction across the board (down to lows of fully decohered states).  The results from figure \ref{Fig: QFT Continuous} confirm the impact of noise to be of the former case, showing that on average the measured $f_1$ rates for both geometries across the full range of $\phi$ are around 60 to 75\% of their theoretical values.  In terms of NISQ Era algorithm design, this means that quantum algorithms which may rely on low probabilities are still viable, whereas noise of the latter case would be far more detrimental.

	\section{Conclusions}
	
	The experimental results in this paper have showcased various qualities of IBM's 20-qubit chip Poughkeepsie.  In analyzing these results, it is important to keep in context the steadily improving technology of quantum computers, specifically superconducting qubits in this case.  In the coming years, it is reasonable to expect quantities such as $T_1$ \& $T_2$ coherence times and gate fidelities to continually improve.  In anticipation of better qubits however, the results from this study demonstrate inherent properties in algorithm design which go beyond qubit quality. 
	
	In testing the CNOT chains across all 20 qubits, the difference in fidelity rates between using $X$ gates versus CNOT gates for resetting ancilla qubits was very pronounced.  This in turn demonstrates the potential for improving algorithm design when working with qubit geometries of limited connectivity, where knowledge of where and when qubits contain superposition in a circuit can be used to optimize ancilla qubit resetting.  Additionally, the study of the various qubit geometries and their performance in implementing the CCNOT and QFT$^{\dagger}$ operations further showcased the challenges of circuit design with limited connectivity.  When determining the best geometry of computational and ancilla qubits for implementing an algorithm, our results demonstrated pros and cons of various configurations, ultimately showing that for certain circuits the use of ancilla qubits can be used to potentially reduce total gate counts.

	\section*{Acknowledgments}
	
	We gratefully acknowledge support from the National Research Council Associateship Program.  Special thanks to Dan Campbell for his numerous insightful talks throughout this project. We would also like to thank the IBM Quantum Experience team and all of their support.  Any opinions, findings, conclusions or recommendations expressed in this material are those of the author(s) and do not necessarily reflect the views of AFRL.
	
	\subsection{Data Availability}
	
	The data that support the findings of this study are available from the corresponding author (dkoch.afrl@gmail.com) upon reasonable request.
	
	\pagebreak


\begin{thebibliography}{99}
		\bibitem{nisq} J.~Preskill, Quantum, volume 2, page 79 (2006)
		\bibitem{shor} P.~Shor, SIAM J. Sci. Statist. Comput. \textbf{26} 1484 (1997)
		\bibitem{grover} L.~K.~Grover, Proceedings of the 28th Annual ACM Symposium on the Theory of Computing (1996)
		\bibitem{noise1} A.~A.~Clerk, M.~H.~Devoret, S.~M.~Girvin, F.~Marquardt and R.~J.~Schoelkopf, Rev. Mod. Phys. \textbf{82}, 1155 (2010).
		\bibitem{noise2} P.~V.~Klimov et. all, Phys. Rev. Lett. \textbf{121}, 090502 (2018)
		\bibitem{noise3} P.~Krantz, M.~Kjaergaard, F.~Yan, T.~P.~Orlando, S.~Gustavsson and W.~D.~Oliver, Applied Phys. Rev. \textbf{6}, 021318 (2019)
		\bibitem{noise4} M.~Kjaergaard, M.~E.~Schwartz, J.~Braumüller, P.~Krantz, J.~I-J.~Wang, S.~Gustavsson and William D. Oliver, arXiv: 1905.13641 (2019)
		\bibitem{err1} K.~Temme, S.~Bravyi, and J.~M.~Gambetta Phys. Rev. Lett. \textbf{119}, 180509
		\bibitem{err2} R.~Harper and S.~T.~Flammia, Phys. Rev. Lett. \textbf{122}, 080504 (2019)
		\bibitem{err3} P.~Murali, D.~C.~McKay, M.~Martonosi and A.~Javadi-Abhari, arXiv: 2001.02826 (2020)
		\bibitem{err_corr1} P.~Shor, Phys, Rev. A \textbf{52}, R2493 (1995).
		\bibitem{err_corr2} A.~Calderbank and P.~Shor, Phys. Rev. A \textbf{54}, 1098 (1996).
		\bibitem{err_corr3} A.~Steane, Phys. Rev. Lett. \textbf{77}, 793 (1996).
		\bibitem{ibmq} IBM 20-Qubit Poughkeepsie Architecture, https://quantum-computing.ibm.com. Accessed Sep. - Dec. 2019
		\bibitem{google} Google's Cirq Quantum Computing, https://github.com/quantumlib/Cirq (2019)
		\bibitem{rigetti} Rigetti's Pyquil Quantum Computing, https://www.rigetti.com/ (2019)
		\bibitem{microsoft} Microsoft's Azure Quantum Computing, https://azure.microsoft.com/en-us/services/quantum/ (2019)
		\bibitem{tutorials1} D.~Koch, L.~Wessing, P.~M.~Alsing, arXiv:1903.04359 (2019)
		\bibitem{tutorials2} D.~Koch, S.~Patel, L.~Wessing, P.~M.~Alsing, arXiv:2008.10647  (2020)
		\bibitem{T2_1} E.~L.~Hahn Phys. Rev. \textbf{80} (4), 580–594 (1950)
		\bibitem{T2_2} N.~F.~Ramsey, Phys. Rev. \textbf{78} (6), 695–699. (1950)
		\bibitem{T1_T2} X.~R.~Wang, Y.~S.~Zheng, and S.~Yin, Phys. Rev. B \textbf{72}, 121303(R) (2005)
		\bibitem{qft} D.~Coppersmith, arXiv: 0201067 (1994)
		\bibitem{toffoli} T.~Toffoli (1980), J.~W.~de Bakker and J.~van Leeuwen (eds.) \textit{Automata, Languages and Programming}, pg.632 Springer: New York (1988)
		\bibitem{ibmq1}N.~M.~Linke, D.~Maslov, M.~Roetteler, S.~Debnath, C.~Figgatt, K.~A.~Landsman, K.~Wright, and C.~Monroe, Proceedings of the National Academy of Sciences \textbf{114}, 13 (2017)
		\bibitem{ibmq2}S.~S.~Tannu and M.~K.~Qureshi arXiv:quant-ph/1805.10224 (2018)
		\bibitem{ibmq3}P.~J.~Coles, S.~Eidenbenz et al., arXiv:1804.03719 (2018)
		\bibitem{ibm_exp1} R.~Balu, D.~Castillo, and G.~Siopsis, Quantum Science and Technology, IOP Publishing (2018)		
		\bibitem{ibm_exp2} A.~May, L.~Schlieper and J.~Schwinger, arXiv: 1910.00802 (2019)
		\bibitem{ibm_exp3} H.~Mohammadbagherpoor, Y-H.~Oh, P.~Dreher, A.~Singh, X.~Yu and A.~J.~Rindos, arXiv: 1910.11696 (2019)
		\bibitem{ibm_exp4} E.~Buksman, A.~L.~Fonseca de Oliveira and C.~Allende, arXiv: 1912.07486  (2019)
		\bibitem{ibm_exp5} 	P.~M.~Q.~Cruz, G.~ Catarina, R.~ Gautier and J.~ Fernández-Rossier, Quantum Sci. Technol. \textbf{5} 044005 (2020)
		\bibitem{book1} M.~O.~Scully and M.~S.~Zubairy, \textit{Quantum Optics}, Cambridge University Press (1997)
		\bibitem{toffoli2} A.~Barenco, C.~H.~Bennett, R.~Cleve, D.~P.~DiVincenzo, N.~Margolus, P.~Shor, T.~Sleator, J.~A.~Smolin, and H.~Weinfurter, Phys. Rev. A. \textbf{52}, 3457 (1995)
		\bibitem{shor2} T.~Häner, M.~Roetteler, and K.~M.~Svore, Quantum Information and Computation, Vol. \textbf{17}, No. 7 (2017)
		\bibitem{qpe} A.~Y.~Kitaev, arXiv: 9511026 (1995)
		\bibitem{qaoa}	E.~Farhi, J.~Goldstone, and S.~Gutmann, arXiv:1411.4028 (2014)
		\bibitem{book2} M.~Nielsen and I.~Chuang, \textit{Quantum Computation and Quantum Information}, Cambridge University Press (2000)
	\end{thebibliography}
\end{document}